\documentclass[twocolumn,pre]{revtex4-1}

\usepackage[utf8]{inputenc}
\usepackage{amssymb,amsthm,amsmath,amsbsy}
\usepackage{graphicx}
\usepackage{bm}

\begin{document}
\title{Fluctuation spectra of large random dynamical systems reveal \\hidden structure in ecological networks}

\author{Yvonne Krumbeck$^1$}

\author{Qian Yang$^2$}

\author{George W. A. Constable$^3$}

\author{Tim Rogers$^{1\dag}$}


\begin{abstract} 
\noindent\textit{$^1$Centre for Networks and Collective Behaviour, Department of Mathematical Sciences, \\University of Bath, Bath, BA2 7AY, UK\\ $^2$Beijing Institute of Radiation Medicine, Beijing 100850, P. R. China\\$^3$ Department of Mathematics, University of York, Heslington, York, YO10 5D2D, UK\\$^\dag$ t.c.rogers@bath.ac.uk}
\begin{center}\textbf{ABSTRACT}\end{center}

Understanding the relationship between complexity and stability in large dynamical systems ---such as ecosystems--- remains a key open question in complexity theory which has inspired a rich body of work developed over more than fifty years. 
The vast majority of this theory addresses asymptotic linear stability around equilibrium points, but the idea of `stability' in fact has other uses in the empirical ecological literature. 
The important notion of {`temporal stability'} describes the character of fluctuations in population dynamics, driven by intrinsic or extrinsic noise. Here we apply tools from random matrix theory to the problem of temporal stability, deriving analytical predictions for the fluctuation spectra of complex ecological networks.  We show that different network structures leave distinct signatures in the spectrum of fluctuations, and demonstrate the application of our theory to the analysis ecological timeseries data of plankton abundances. 
\end{abstract}

\maketitle

\section*{Introduction}

``Will a large complex system be stable?" asks the title of Robert May's seminal 1972 paper~\cite{may_will_1972} that threw fuel on the fire of the complexity-stability debate and popularised the use of random matrix theory (RMT) in theoretical ecology. 
At first sight, answering this question with mathematics seems impossible. The huge number of interactions in real-world ecosystems hampers any attempt to create a precisely calibrated model, as the challenge of measuring all necessary parameters seems insurmountable. 
What May pointed out was that it might in fact not be necessary to know exact parameter values; knowledge of their statistical distribution could be sufficient. Combining the random model ecosystems proposed by Gardner and Ashby~\cite{gardner_connectance_1970}, with results of Ginibre \cite{ginibre1965statistical} in RMT, May showed how complexity ---measured in terms of the number of species and the connectance of their interaction network--- could decrease ecosystem stability. 

Although modelling ecosystems as using random community matrices has been criticised~\cite{james_constructing_2015,jacquet2016no} ---with some arguing that these serve best as a null model for ecosystem structure~\cite{jr_unified_2010}--- this growing field has continued to provide insights into the mechanisms that promote ecosystem stability.
For instance, Allesina and Tang~\cite{allesina_stability_2012, allesina_stabilitycomplexity_2015} generalised the community matrix model to account for different interaction types, elucidating the important stabilising role of predator-prey interactions. We now have quite a detailed view on the extent to which high-level ecosystem information (such as trophic \cite{allesina2015predicting} or community \cite{grilli2016modularity} structures) can be incorporated into the RMT framework to give more accurate predictions of the stability boundary. 

The notion of stability referred to by May and these later works is that of asymptotic linear stability of an equilibrium point.  While this definition is a natural mathematical choice, it belies the rich array of interpretations of {`ecological stability'} present in the empirical ecological literature~\cite{donohue_2016}.
In order to make clear the differences between these interpretations, Grimm and Wissel~\cite{grimm_babel_1997} created an inventory for different types of stability measures used in ecology. Some of these in particular are more attuned to the measures favoured by empirical  ecologists.
One such measure is {`temporal stability'}, often described as the constancy of ecological variables relative to their mean, which is commonly used as an indicator for ecological stability~\cite{levins_coexistence_1979, ives_stability_1999, lehman_biodiversity_2000, tilman_biodiversity_2006, jiang_different_2009, loreau_species_2008, campbell_experimental_2011, donohue_dimensionality_2013}. In \cite{suweis2015effect}, Suweis {\it et al.} propose to study the attenuation of perturbations as they propagate through ecological networks, introducing measures of reactivity and localization. Taking a different approach, recently Arnoldi et al.~\cite{arnoldi_resilience_2016} employed the term {`variability'} to describe the inverse of temporal stability in a random community matrix model. In that work, they consider the scale of response to persistent external (environmental) noise applied to an ecosystem. While this is an important and useful measure, it does not capture anything of the temporal characteristics of fluctuations in ecosystems, which can drive a system away from equilibrium, and thus are important precursors to linear and nonlinear instabilities \cite{wiesenfeld1985noisy}. 

\begin{figure}[t]
    \includegraphics[width=0.48\textwidth,trim=20 12 10 0,clip=true]{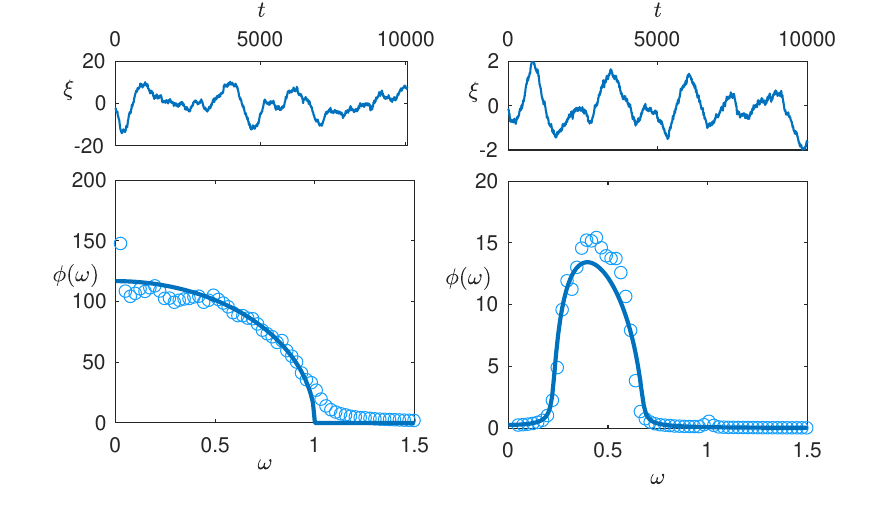}
    \caption{Fluctuation spectra in predator-prey systems. Left: a mixed community of species with randomly assigned predation relationships. Right: a model ecosystem with two trophic levels--- 200 predators and 800 prey species. In both cases fluctuations are illustrated via a typical single-species time series ($\xi(t)$, upper) and the mean power spectra ($\phi(\omega)$, lower), where circles are simulation results, and solid lines our theory. Parameters are: $Nx=200, N_y=800, c_x=20, c_y=5, \alpha=10, b=1, d=1$; full details of all simulations are found in the Methods.}
    \label{fig1}
\end{figure}

In this paper, we seek to bridge empirical and theoretical measures of stability by developing a theoretical framework for the analysis of temporal stability of ecosystems. Our key object of study is the {`power spectral density'}, a statistical measure that captures the frequency and amplitude of noisy fluctuations in time series (see Fig.~1 for examples). The relationship between such power spectra and temporal notions of ecological stability is multifaceted. Particular points of interest are the height of spectrum, which gives information about the magnitude of stochastic fluctuations, the locations of non-zero peaks corresponding to quasi-cyclic signals, or a peak at zero indicating baseline wander.  Moreover, the Fourier transform of the spectrum yields the autocorrelation structure of the stochastic trajectories. We provide a brief guide to these concepts in the {Methods section Interpreting the Power Spectral Density in the Context of Temporal Stability}, though in the main text we will refrain from ascribing overly simplified interpretations to power spectra.

Beyond providing a more detailed view of temporal stability, an investigation of power spectra yields a number of further advantages. For instance, power spectra are readily computed from empirical data and provide detailed information about intrinsic fluctuations and (via the fluctuation dissipation theorem~\cite{kubo1966fluctuation}) response to external perturbations. Previous theoretical studies of power spectra in low-dimensional systems have yielded important and sometimes surprising results in fields including epidemiology, game theory, and ecology~\cite{alonso2007stochastic,galla2009intrinsic,mckane_predator-prey_2005}. The method is particularly powerful in explaining the emergence of persistent quasi-cyclic oscillations driven by noise. 
Until now, however, a major limitation of this theory has been its restriction to models with very small numbers of interacting elements for which the approach is analytically tractable with existing methods, while the applicability of the theory to larger systems is limited by comparatively slow numerical schemes, and difficulty parameterising large models.
Here, by applying techniques from the statistical physics of complex systems, we demonstrate the possibility of deriving exact analytic formulae for the power spectra of large random and noisy dynamical systems. 

We apply our method to characterise the stochastic fluctuations of species abundances in random Lotka-Volterra type ecosystem models.
As a result, we find that their temporal stability is universally characterised by a few key parameters, including the proportion of predator-prey interactions and the rate of population turnover.
This result is a temporal analogue of the famous Winger semi-circle law for random matrix eigenvalue distributions \cite{wigner1958distribution} and points to the wide applicability of the theory we develop. Importantly, the universal character of the power spectrum we derive is independent of the choice of random variables in the model, and only depends on the aggregate properties we identify. 

Just as May's RMT calculations are open to generalisations and refinements, so too is our approach to temporal stability. 
We illustrate this flexibility of the theory by incorporating trophic structure to our ecosystem models.
Subsequently, we discover a distinct signature of this type of structure: the confinement of fluctuations to a fixed band of frequencies.
Taken together, these results raise the exciting prospect of being able to draw conclusions about the internal structure of an ecosystem through the analysis of its fluctuations.

The paper is structured as follows. 
First we demonstrate how to compute the mean power spectral density of a large random Lotka-Volterra system in {section Interaction types determine fluctuation spectra}, showing how different dominant interaction types result in distinct fluctuation power spectra. We then show how to compute the spectrum for an individual species within the large random ecosystem system in {section Species fluctuations exhibit strong heterogeneity}, and further generalise the method in {section Trophic structure induces fluctuation frequency gap} to consider bipartite interaction networks, showing how a two-level trophic system can leave a distinct fingerprint in the power spectrum of an ecosystem. 
Readers interested in the potential of our results as a tool for analysis of real time series data may wish to jump to {section Confronting RMT in theoretical ecology with time series data} which provides a proof of concept in this direction. Here we explore an ecological time series dataset of plankton abundances, showing how our results provide a technique to infer the structural details of real ecosystems. 
Full derivations of our analytic results are provided in the Methods section, along with detailed descriptions of the models we use for demonstration throughout this paper.


\section*{Results}

\subsection*{Interaction types determine fluctuation spectra}\label{sec_interaction_types}

Our approach enables the computation of the power spectral density of fluctuations in large random systems of a very general class; a full and detailed derivation is given in the Methods. In the case of ecosystem stability, the dynamical system in question is that describing the interactions of different species. Many modelling choices are possible in this context. For clarity we will focus here on an established modelling paradigm --- large Lotka-Volterra type ecosystems ---  and explore the extent to which the nature of the species interactions affects the shape of the fluctuation spectrum. 

Following classic models of ecosystem dynamics, we consider $N$ species occupying a domain of size $V$, writing $x_i(t)$ for the density of individuals of species $i$ at time $t$. For large but finite $V$, standard techniques (see Methods) allow us to describe the change of the species densities by a set of stochastic differential equations (SDEs) :
\begin{equation}
    \frac{dx_i}{dt} = x_i \left( b_i +\sum_{j}^N \alpha_{ij}x_j \right) + \frac{1}{\sqrt{V}} \eta_i(t).
    \label{eq_HLV_sde_main}
\end{equation}
Here, the coefficients $\alpha_{ij}$ for $i \ne j$ describe the interaction between species $i$ and $j$, and the $\eta_i(t)$ are Gaussian noise term with correlations ${\langle\eta_i(t)\eta_j(t')\rangle=\delta(t-t')B_{ij}(\bm{x})}$.

We parameterise the model as follows. For simplicity (and to isolate the effect of interaction types) we model each species as having the same birth rate $b_i\equiv b$ and density dependent mortality rate $\alpha_{ii}\equiv-b$. The other interaction coefficients $\alpha_{ij}$ are chosen at random so that (i) each species interacts with an average of $c$ others (for each possible interaction we include it with probability $c/N$, independent of all others), (ii) interactions have mean strength $\mathbb{E}|\alpha_{ij}|=\mu$ and second moment $\mathbb{E}\alpha_{ij}^2=\sigma^2$, (iii) the correlation is controlled by the symmetry parameter $\gamma=\mathbb{E}[\alpha_{ij}\alpha_{ji}]/\sigma^2\in[-1,1]$. Crucially, the full details of the distribution of the parameters $\alpha_{ij}$ are not required, thanks to the universality of property of large random matrices \cite{tao2008random,tao2010random}.

In the methods we show how these rates can be derived from a simple model of pairwise species interactions which can be mutualistic, competitive, or predatory. The frequency of predator-prey type interactions is tied to the symmetry parameter $\gamma$. At $\gamma=-1$ all interactions are of the predator-prey type ($\alpha_{ij}=-\alpha_{ji}$), at $\gamma=1$ only purely mutualistic ($\alpha_{ij}=\alpha_{ji}>0$) or competitive ($\alpha_{ij}=\alpha_{ji}<0$) are present, and between these extremes there is a random mix of interaction types. 

With this choice of (random) parameters, each species density will fluctuate around the scaled carrying capacity $x^*_i=1$, which, following \cite{allesina_stabilitycomplexity_2015}, is stable provided $b>\sqrt{c\sigma^2}(1+\gamma)$ (we refer to~\cite{stone_feasibility_2018, gibbs_effect_2018} for stability and feasibility of equilibrium states with heterogeneous species abundance distributions). Around this fixed point, species in the stochastic system in Eq.~(\ref{eq_HLV_sde_main}) will exhibit approximately linear fluctuations $\xi_i$, described by an Ornstein-Uhlenbeck process of the form 
\begin{equation}
\frac{d\bm{\xi}}{dt}=\bm{A}\bm{\xi}+\bm{\zeta}(t)\,.
\label{eq2}
\end{equation}
Here $\bm{A}$ is the Jacobian of Eq.~(\ref{eq_HLV_sde_main}), known as the community matrix in the context of theoretical ecology, and $\bm{\zeta}$ is an $N$-vector of Gaussian white noise with correlation matrix $\bm{B}=\bm{B}(\bm{x}^*)$. We assume that the equilibrium point at $\bm{x}=\bm{x}^{*}$ is linearly asymptotically stable (i.e. stable in the mathematical sense described by May \cite{may_will_1972}) and now proceed to investigate its temporal stability as characterised by the power spectra (see Methods {section Interpreting the Power Spectral Density in the Context of Temporal Stability}).

The power spectral density of fluctuations $\boldsymbol{\Phi}(\omega)$  is defined as the Fourier transform of the covariance ${\mathbb{E}[ \boldsymbol{\xi}(t)\boldsymbol{\xi}(t+\tau)^T ]}$. For multivariate Ornstein-Uhlenbeck processes one can show (see e.g. \cite{gardiner_stochastic_2009}) that 
\begin{equation}
\begin{split}
    \boldsymbol{\Phi}(\omega) &:= \int_{-\infty}^{\infty} \mathrm{e}^{-\text{i}\omega\tau}\mathbb{E}[ \boldsymbol{\xi}(t)\boldsymbol{\xi}(t+\tau) ] d\tau\\
    &=(\boldsymbol{A}-i\omega\boldsymbol{I})^{-1} \boldsymbol{B} (\boldsymbol{A}^T+i\omega\boldsymbol{I})^{-1}.
    \label{eq_powerspectrum_definition_main}
    \end{split}
\end{equation}

In the Methods we show how to apply random matrix theory techniques to compute the power spectral density, via a complex Gaussian integral representation of the above matrix equation. The approach provides a general framework for computing the fluctuations in large systems specified by random matrices $\bm{A}$ and $\bm{B}$. We derive an expression for the mean-field power spectral density $\phi(\omega)=\mathbb{E}[\bm{\Phi}_{ii}]$ in terms of the resolvent function $r$. Specifically, 
\begin{equation}
    \phi = |r|^2\frac{\mathbb{E}[B_{ii}] +2\mathrm{Re}(r)c\mathbb{E}[A_{ij}B_{ij}]}{1-|r|^2c\mathbb{E}[A_{ij}^2]} \,,
    \label{eq_powerspec_meanfield}
\end{equation}
where expectation is taken only over the non-zero entries of $\bm{A}$ and $\bm{B}$, and $r\in\mathbb{C}$ solves the self-consistent equation
\begin{equation}
\frac{1}{r} = -\mathbb{E}[A_{ii}] +i\omega -rc\mathbb{E}[A_{ij}A_{ji}] \,.
\end{equation}

This result holds for general random matrix models in which interaction parameters are drawn from the same distribution for all species pairs (and we later show how the method can be extend for other model types with species-specific parameters using a single-defect approximation or partitioned networks). For the present case of our Lotka-Volterra model, the community matrix coincides with the interaction matrix (that is $A_{ij}=\alpha_{ij}$). In the methods we derive the rules $\mathbb{E}[B_{ii}]=2b+c\mu$, $\mathbb{E}[A_{ij}B_{ij}]=0$ for the statistics of the noise correlation matrix. To get a sense for the information contained in Eq.~(\ref{eq_powerspec_meanfield}), we explore the result for several cases with varying interaction structures. 

\begin{figure}[t]
    \centering
    \includegraphics[width=0.5\textwidth, clip=20 0 0 0]{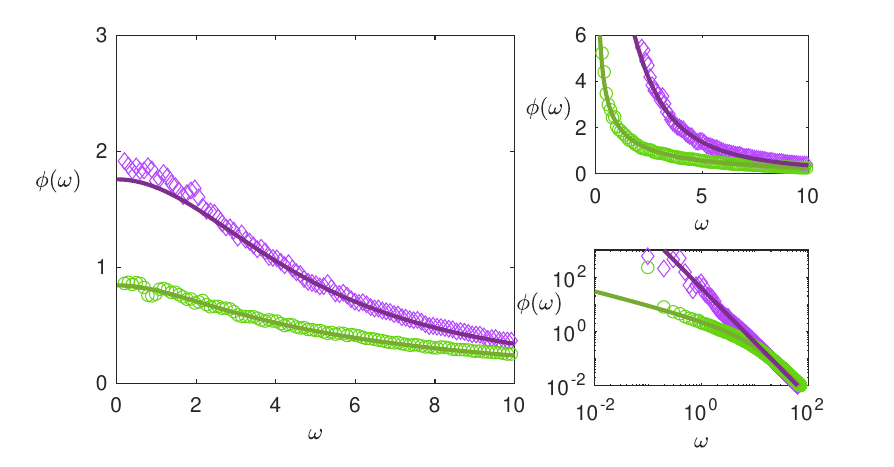}    
	\caption{{Fluctuation spectra in systems with lower predator-prey densisty.} Left: stable systems with $\gamma=0$ (purple {diamonds}) and $\gamma=1$ (green {circles}) have similar fluctuation spectra, with a higher proportion of predator prey interactions associated with higher over all excitation. Right upper: near instability a pole emerges at $\omega=0$, corresponding to baseline drift in marginally stable systems. Right lower: In log-log axes we see the different nature of the pole for mixed ($\gamma=0$, purple {diamonds}) communities compared with those with only symmetric interactions ($\gamma=1$, green {}circles). For $\gamma=-1$, see Fig.~\ref{fig1}, left panel. Common parameters are $N=1000, c=50, \sigma^2=0.5$.}
    \label{fig2}
 \end{figure}

First consider a weak interaction limit where the difference between species is rather small, so that $\sigma^2\ll1$. In this case we find a simple Lorentzian spectral density:
\begin{equation}
    \begin{split}
        \phi(\omega) &= \frac{2b +c\mu}{b^2 +\omega^2} +\mathcal{O}(\sigma^2). 
    \end{split}
    \label{eq_HLV_powerspec_weak}
\end{equation}
Fluctuations of this type are indicative of a highly stable system in which the balance of interaction types $\gamma$ has no influence. Next let us consider a limit where the power spectral density shows significant differences depending on the proportion of predator-prey interactions within the community, in particular focussing on ecosystems that are near the stability boundary. 

In the case of a system with predator-prey interactions only, we have $\gamma=-1$ and the ecosystem is stable for all positive birth rates $b$. Expanding in small $b$ we find that fluctuations are of order $1/b$, but are almost completely confined to a low-frequency window. If $\omega^2<4c\sigma^2$ then 
\begin{equation}
    \begin{split}
        \phi(\omega) &= 
            \frac{2b +c\mu}{2c\sigma^2} \left[ \frac{1}{b} \sqrt{4c\sigma^2-\omega^2} -\frac{c\mu}{2b+c\mu}\right] +\mathcal{O}(b), 
    \end{split}
    \label{eq_HLV_powerspec_predprey_main}
\end{equation}
with an order $1/\omega$ tail outside this range. Note that Eq.~(\ref{eq_HLV_powerspec_predprey_main}) has the shape of a quarter-circle, to be viewed as a natural counterpart to the Wigner semi-circle law in classical random matrix theory \cite{wigner1958distribution}. The result is illustrated in Fig.~\ref{fig1}, left panel.

For a random mixture of interaction types with $\gamma=0$ no approximations are necessary as Eq.~(\ref{eq_powerspec_meanfield}) simplifies to 
\begin{equation}
    \begin{split}
        \phi(\omega) &= \frac{2b +c\mu}{b^2 -c\sigma^2 +\omega^2}.
    \end{split}
    \label{eq_HLV_powerspec_mix_main}
\end{equation}
The stability boundary here is given by $b^2=c\sigma^2$. The above result therefore implies the emergence of a $1/\omega^2$ divergence in the power spectrum at low frequencies when such a system is close to instability (see Fig.~\ref{fig2}). 

When only mutualistic or competitive interactions are present (i.e. $\gamma=+1$), the full solution to Eq.~(\ref{eq_HLV_powerspec_predprey_main}) in this case is complicated, but for stable systems appears qualitatively similar to the result Eq.~(\ref{eq_HLV_powerspec_mix_main}) above. Near the stability boundary, however, we find another behaviour. When $b^2=4c\sigma^2$, we find 
\begin{equation}
        \phi(\omega) = \frac{\sqrt{2}(4\sqrt{c\sigma^2}+c\mu)}{\sqrt{c\sigma^2(\sqrt{16c\sigma^2\omega^2+\omega^4}-\omega^2})}-\frac{(4\sqrt{c\sigma^2}+c\mu)}{2c\sigma^2}.      
    \label{eq_HLV_powerspec_symmetric_main}
\end{equation}
In contrast to the previous case, this power spectrum exhibits a pole of order $1/\sqrt{\omega}$ at low-frequency, followed by a $1/\omega^2$ tail at high frequency (see Fig.~\ref{fig2}).  

Between these results, we are able to see how the proportion of predator-prey interactions in an ecosystem leaves a signature in the fluctuation spectrum. When predator-prey interactions are dominant, the shape of the spectrum is pulled towards a quarter circle law (Fig.~1, left panel); when they are rare, the low-frequency pole near instability changes its character (Fig.~2). 

\subsection*{Species fluctuations exhibit strong heterogeneity}\label{sec_SDA}

So far, we have considered only the mean power spectral density of fluctuations. The cavity method technology employed in the derivation of Eq.~(\ref{eq_powerspec_meanfield}) can also yield detailed information about the fluctuation spectra of individual species in an ecosystem model. Suppose one is interested in a focal species $i$, and has data on the type and strength of interactions this species has with others in its ecosystem, as well as an estimate of the large scale ecosystem parameters such appearing in Eq.~(\ref{eq_powerspec_meanfield}). It is possible to make use of this data in a {`single defect approximation'} (SDA) scheme in which one considers the fluctuations of species $i$ when embedded in a large unknown ecosystem. 

In the Methods we show how to derive an SDA approximation $\phi_i^{\text{SDA}}$ to the spectral density of fluctuations for species $i$, given by the expression 
\begin{equation}
    \phi_i^{\text{SDA}}= \frac{\phi^{\text{MF}}\sum_{i\sim j}A_{ij}^2 + 2\mathrm{Re}(r^{\text{MF}})\sum_{i\sim j}A_{ij}B_{ij}+ B_{ii}}{|A_{ii}+i\omega+\bar{r}^{\text{MF}}\sum_{i\sim j}A_{ij}A_{ji}|^2} \,,
    \label{eq_powerspec_recursion_SDA_main}
\end{equation}
where $\phi^{\text{MF}}$ and $r^{\text{MF}}$ are the mean-field power spectrum and resolvent obeying Eq.~(\ref{eq_powerspec_meanfield}). 

\begin{figure}[t]
    \centering
    \includegraphics[width=0.5\textwidth, trim=110 300 100 300, clip=true]{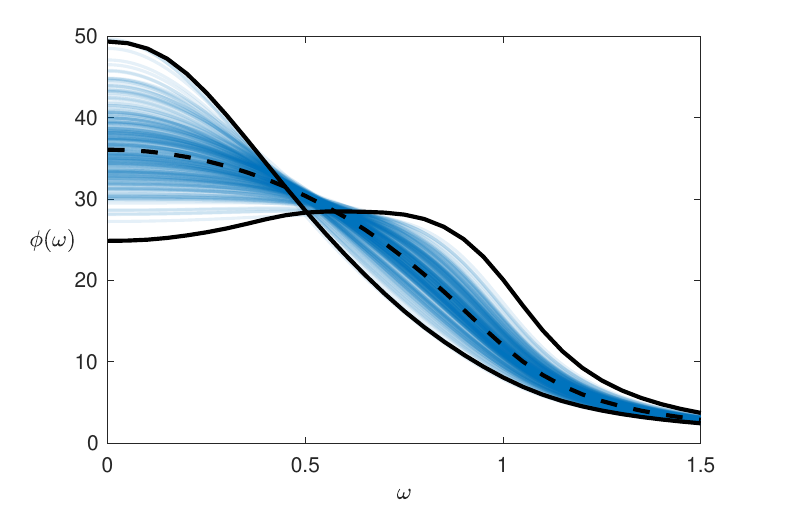}
    \caption{Heterogeneity in species fluctuations. Thin lines show power spectral densities for individual species in a predator-prey ecosystem model, computed using the single-defect approximation of Eq.~(\ref{eq_powerspec_recursion_SDA_main}). Thick lines show the power spectral density for comparison the two species with extremal fluctuations at low frequency, computed directly via Eq.~(\ref{eq_powerspectrum_definition}); the dash line shows the mean power spectral density.
Parameters are: $N=500, c=20, \gamma=-1, \sigma^2=1/4/c, b=0.2$. }
    \label{fig_SDA_powerspec}
\end{figure}

In Fig.~\ref{fig_SDA_powerspec} we compare the average power spectral density of all species with the spectra of individual species as computed directly and via the SDA approximation. We immediately notice that the mean-field power spectral density is often not representative of individual species, which show surprisingly strong heterogeneity in their fluctuation spectra. Another interesting feature of these results is the presence of peaks in the power spectral density away from zero for some species --- this implies quasi-periodic fluctuations in these populations that are not observed in the ecosystem as a whole.  

Finally we observe the curious feature that (for this model at least) the total power of fluctuations appears approximately conserved, meaning that those species which do not have large fluctuations at low frequencies are the same as those with unusually large fluctuations at higher frequencies. At present we do not have an intuitive explanation for this behaviour, highlighting the richness of non-obvious information present in these complex power spectral densities.

\subsection*{Trophic structure induces fluctuation frequency gap}\label{sec_trophic}

In the above investigations, we have employed a simple ecosystem model in which species interactions are assigned completely at random. In the past fifty years of research into random matrix ecosystem models, far more sophisticated and realistic models have been developed. Let us now illustrate how our methods may be applied to more detailed models using the example of ecosystems with explicit trophic structure. Here we focus on a bipartite predator-prey network as an example.

Consider a large model ecosystem composed of $N_x$ predator species and $N_y$ prey species, writing $x_i$ and $y_j$ for the density of predator species $i$ and prey species $j$, respectively. With no prey-prey or predator-predator interactions, the interaction structure is bipartite. Each predator species has an extrinsic death rate $d$ and depends upon the consumption of prey for reproduction. This consumption may come from a selection of $c_x$ different prey species for each predator, with $R_{ij} > 0$ giving the predation rate of predator $i$ on prey $j$. Conversely, each prey has birth rate $b$, but is hunted by $c_y$ predators, where $N_xc_x=N_yc_y$. The SDEs for the predator and prey densities are given by
\begin{equation}
    \begin{split}
        \frac{dx_i}{dt} &= x_i \left(-d -x_i + \sum_j^{N_y} R_{ij}y_j \right) +\frac{1}{\sqrt{V}}\eta_i(t), \\
        \frac{dy_j}{dt} &= y_j \left(b -y_j - \sum_i^{N_x} R_{ji}x_i \right) +\frac{1}{\sqrt{V}}\eta_j(t), \\
    \end{split}
    \label{eq_bipartite_sde_main}
\end{equation}
where $\eta_{i,j}(t)$ are Gaussian noise with $\langle\eta_i(t),\eta_j(t')\rangle=\delta_{ij}(t-t')B_{ij}(\bm{x},\bm{y})$. In the Methods we show how these equations (and the specific form of $B_{ij}$) are derived from an individual-based model. 

This model has an equilibrium state $(\bm{x}^*,\bm{y}^*)$, around which linear-order fluctuations will occur, analogously to Eq.~(\ref{eq2}) above. We compute a community matrix of the form
\begin{equation}
    \boldsymbol{A} = \left(\begin{matrix}
        -x^*\boldsymbol{I} & x^*\boldsymbol{R} \\
        -y^*\boldsymbol{R}^T & -y^*\boldsymbol{I}
    \end{matrix}\right),
\end{equation} 
where the first ${i=1,\dots,N_x}$ rows and columns represent the predator species, and the remaining ${j=N_x+1,\dots,N_x+N_y}$ rows and columns correspond to the prey species. The noise matrix is derived from the underlying individual-based model (see Methods) and given by
\begin{equation}
    \boldsymbol{B} = \left(\begin{matrix}
        2x^*(x^*+d)\boldsymbol{I} & -x^*y^*\boldsymbol{R} \\
        -x^*y^*\boldsymbol{R}^T & 2y^*b\boldsymbol{I}
    \end{matrix}\right).
\end{equation} 

\begin{figure}[t]
    \centering
    \includegraphics[width=0.47\textwidth]{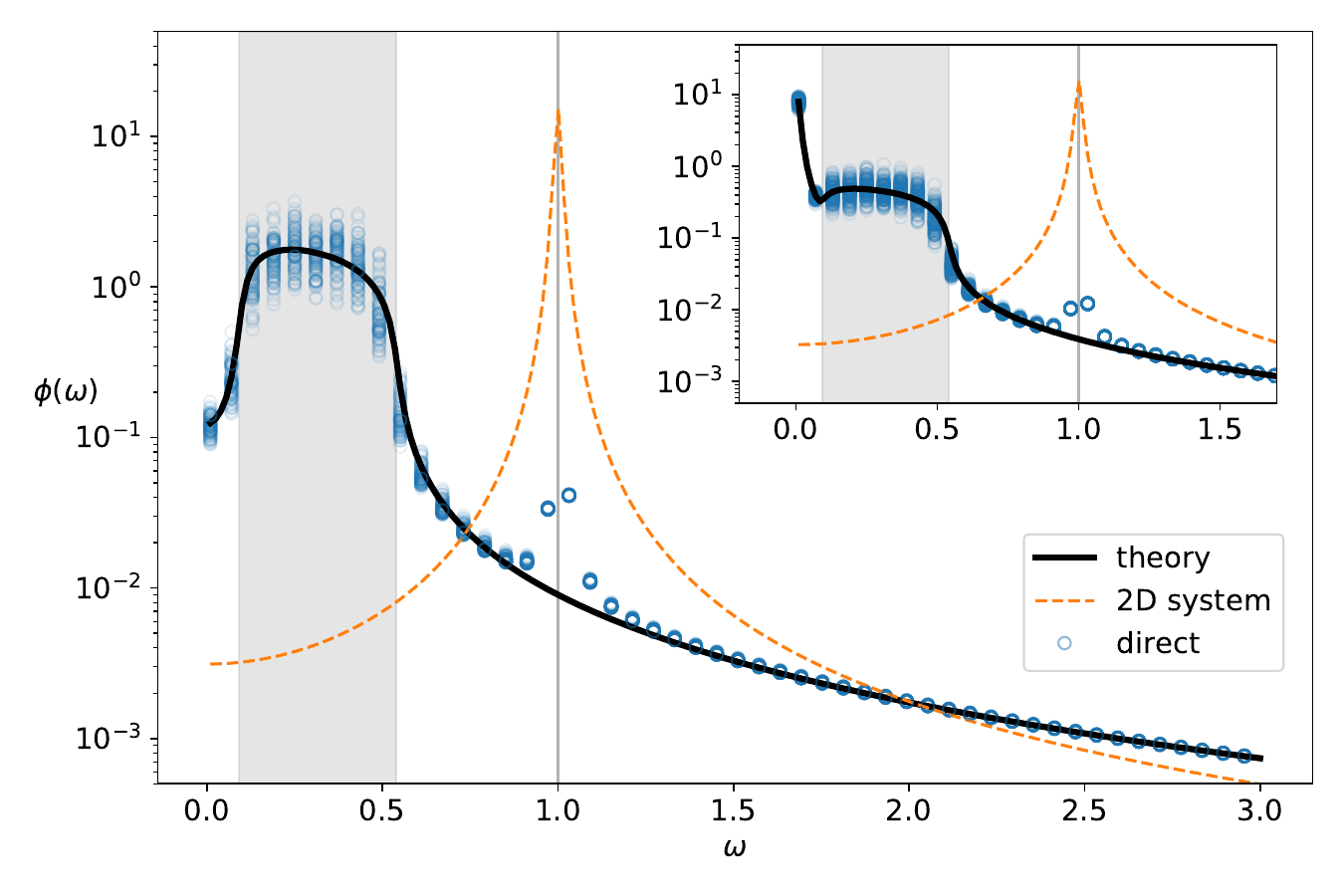}
    \caption{{Fluctuation spectra of bipartite systems.} The exact power spectral densities (dots) for predator species in a two trophic level model {are} computed numerically from Eq.~(\ref{eq_powerspectrum_definition}), and the corresponding mean power spectral density (solid line) {are} obtained by solving Eq.~(\ref{eq_powerspec_recursion_BPP}) (inset shows the prey species).
    Excited frequencies are confined to a band (shaded) between critical frequences given in Eq.~(\ref{eq_BPP_band_gap}).   
    The order $1/N$ peak at $\omega=1$ in the simulation result relates to the high-level bipartite structure. Its location is predicted by a corresponding 2D system, also shown here for comparison (dashed line). Parameters: $N_x=100, N_y=200, c_x=20, c_y=10, \alpha=5, b=1, d=1$.
    }
    \label{fig_powerspec_BPP_log}
\end{figure}

In the Methods we develop a general approach to computing the power spectral density of large random systems with bipartite structure such as this. The method requires explicitly keeping track of the contributions associated to each species group and their interactions. In the mean-field, this approach delivers a set of equations (\ref{eq_powerspec_recursion_BPP}) to be solved for the mean contributions to the resolvent  $r_x,r_y$, and to the power spectrum, $\phi_x,\phi_y$. Fig~\ref{fig_powerspec_BPP_log} shows the shape of the power spectrum for predator and prey species in this bipartite ecosystem. Surprisingly, we find that fluctuations are mainly confined to a narrow window of frequencies, with a gap in excited frequencies around zero. Examination of the system in Eq.~(\ref{eq_powerspec_recursion_BPP}) allows us to determine the window of excited frequencies to be bounded by the critical frequencies
\begin{equation}
        \omega_\pm = \sqrt{bd \left(\frac{1}{c_x}+\frac{1}{c_y} \pm\frac{2}{\sqrt{c_xc_y}} \right)}.
            \label{eq_BPP_band_gap}
\end{equation}

The contrast between the power spectral density of this two-trophic-level model to that of a mixed ecosystem with predator-prey interactions was illustrated in Fig.~\ref{fig1}. In Fig.~\ref{fig_powerspec_BPP_log} we show the spectrum in more detail, highlighting the band of excited frequencies predicted by Eq.~(\ref{eq_BPP_band_gap}). In the present context, it means that observed time series will not exhibit baseline wander and can therefore be considered to have a higher long-term temporal stability than the mixed ecosystems explored above (see {section Interpreting the Power Spectral Density in the Context of Temporal Stability}). 

Comparisons between simulations and our analytical results shows another interesting feature: an order $1/N$ disagreement at frequency $\omega=1$, which is outside of the excited range. This can be explained by considering an effective two-species model in which we consider only a single {`average'} predator and prey pair. This 2D system has an eigenvalue pair with unit imaginary part, giving rise to quasi-cycle behaviour as documented in~\cite{mckane_predator-prey_2005}. It is important to note that this contribution is small relative to the rest of the spectral density, meaning that the bulk of fluctuations of a structured ecosystem cannot be inferred from considering a low-dimensional representative model. 

\subsection*{Confronting RMT in theoretical ecology with time series data}\label{sec_data}

Although hugely influential in the field of theoretical ecology over the last $50$ years, traditional work on RMT has so far led to rather limited empirically testable insights. The central issue is that while many ecological considerations can be incorporated in a random matrix model, each leads to a binary outcome; the system is either stable or unstable to small perturbations. Thus testing the predictions of these models demands the time-intensive task of measuring real species interaction networks (which are assumed to be stable) and asking whether they indeed tend to be weakly connected (as suggested by May~\cite{may_will_1972}), have a dominance of predator-prey interactions (as suggested by Allesina and Tang~\cite{allesina_stability_2012}), or satisfy some other prediction of the theory. In contrast, the approach presented in this paper offers the tantalizing prospect of directly linking the ecological RMT framework with comparatively easy-to-obtain time series data.

\begin{figure}[t]
    \centering
    \includegraphics[width=0.5\textwidth, trim=10 20 20 0]{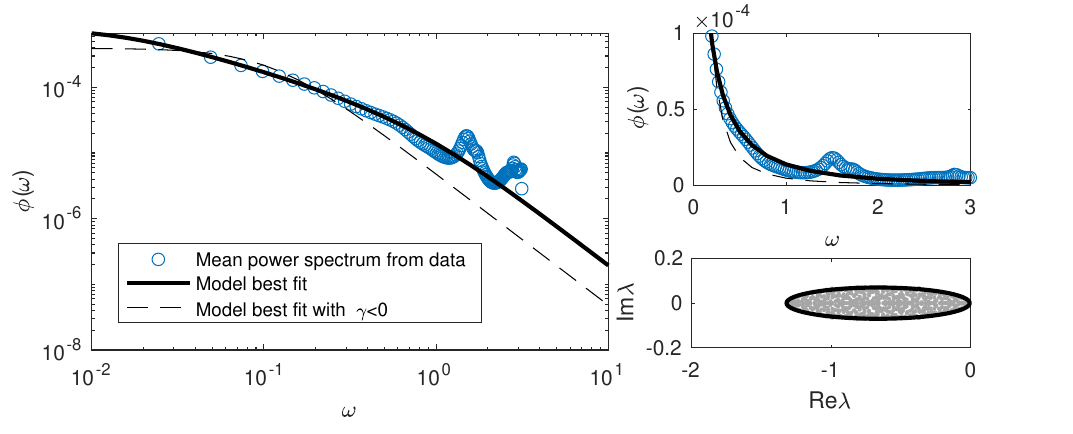}
    \caption{ 
{Fluctuation spectra in plankton species abundance. We plot the power spectral density of data taken} from \cite{plankton_timeseries} (blue {circles}) alongside that of a fitted Lotka-Volterra random ecosystem model (solid line). Fitted parameters are $b=0.6643, c\sigma^2=0.1316, \gamma=0.8078$, data averaged over $n=3$ samples per day. Also shown is the best fit under the restriction $\gamma<0$ (dashed line). The lower right panel shows the spectral boundary inferred from the fit (black ellipse), along with the eigenvalues of a sample random community matrix for illustration. 
    }
    \label{fig_data}
\end{figure}

To trial the use of our methods in the analysis of real ecological data, we have investigated a high-resolution time series dataset for the abundance of coastal plankton species, taken over a period of 88 consecutive days~\cite{plankton_timeseries}. In Figure ~\ref{fig_data} we show the estimated empirical mean power spectrum from the data (circles), compared to that of the best fit Lotka-Volterra random ecosystem model according to our theory. Full details of the data analysis and fitting are given in the Methods. Examination of this fit reveals several qualitative features of the implied ecological interactions. 

First, we note that the best fit value for the interaction symmetry parameter is $\gamma=0.81$, implying an ecosystem in which predator-prey interactions are scarce, and is more likely dominated by competition. Trust in this finding is strengthened by the fact that the fit is quite sensitive to this parameter; the dashed line in Figure ~\ref{fig_data} gives the best fit under the constraint $\gamma<0$, which performs poorly, especially for low frequency. 

Interestingly, when viewed in logarithmic axes (Fig.~\ref{fig_data} main panel), the plankton abundance power spectrum appears to exhibit a similar change of scaling between high and low frequency ranges to that seen in Fig.~\ref{fig2} for the case of symmetric interactions near instability. We can assess the closeness to instability by considering the spectrum inferred from the best fit model, as shown in the lower right panel of Figure~\ref{fig_data}. The rightmost edge ($\lambda_{\max}=\sqrt{c\sigma^2}(1+\gamma)-b\approx-0.0086$) is very close to zero, implying ecosystem dynamics which are close to instability. This feature corresponds to the large peak at zero in the power spectral density, which suggests that low frequency perturbations to the overall species abundances are very slow to relax. 

One feature of the spectrum not reproduced by the simple models considered thus far is the smaller additional peak around $\omega\approx1.5$. This peak has a few possible explanations: a external effect of some sort; possible secondary structure in the ecological interaction network, which could manifest on a system wide scale such as the trophic structure analysed in the previous section; or a feature isolated to a smaller number of more dominant species. A further limitation of the model used here is the assumption of uniform species abundance; in reality, species abundances tend to be distributed log-normally, with few species contributing to the majority of ecosystem biomass. Incorporating such model refinements are well-within the bounds analytical tractability for our approach (see, for instance, \cite{gibbs_effect_2018} and \cite{stone_feasibility_2018}); we hope and expect the theoretical groundwork we have developed here  will pave the way for the investigation of such features in future studies. 

In the above, we have illustrated the use of our methods to infer details of the structure and stability of real ecosystems from time series data, as well as to identify departures from the unstructured assumptions of standard RMT models. Indeed, such departures are present in many real world ecosystems, with important consequences for the validity of any predictions made within the standard RMT framework~\cite{james_constructing_2015}. In contrast, RMT has recently found renewed attention in the field of microbiome research, where it is believed that the key conceit of standard RMT models (that communities are unstructured) holds~\cite{coyte_2015}. However in this field, the spectre of model parameterization again raises its head. 

In \cite{coyte_2015}, a species-interaction network presented in \cite{stein_2013} was used to parameterize an RMT model and show that ecological interactions in the microbiome tended to be weak and non-cooperative. The species interaction network determined in \cite{stein_2013} was itself the result of fitting mouse intestinal microbiome abundances to a deterministic generalized Lotka-Volterra model. However, fully fitting this model required disturbing the mouse microbiota away from its equilibrium state using antibiotics ($S$ fixed point species abundances are insufficient to parameterize an $S\times S$ species interaction network, so data on non-equilibrium transient trajectories were required). While such experimental manipulation may be permissible for studying the microbiota of model organisms such as mice, the ethical issues of such experimentation in humans has raised questions about the informativeness of temporal data for understanding microbial communities such as the human gut microbiota~\cite{xiao_2017}.

In contrast to the approach taken in \cite{coyte_2015}, our methodology requires no external perturbation to a host's microbiome, relying as it does solely on the natural demographic fluctuations present in any finite population. In addition, our approach allows the RMT model itself to be directly parameterized through data, rather than requiring the fitting of an intermediate model.


\section*{Discussion}

In this study we have revisited the complexity--stability question in theoretical ecology with a fresh perspective that develops a random matrix theory approach to temporal stability as captured by the power spectrum of fluctuations. We have applied our techniques to calculate analytic formulae describing the mean power spectra of large Lotka-Volterra ecosystems. We find the fluctuations are described by just a few key parameters: the mean, variance and correlation of entries of the community matrix and noise correlator. We further expanded the method to investigate the role of trophic structures in determining temporal stability, demonstrating the flexibility of the method and usage across a broader range of models. Finally we fitted our model to existing time series data sets, that suggest a majority of competitive or mutualistic interactions within plankton ecosystems. In short, our approach allows us to link the large scale statistical properties of interaction parameters with the emergent fluctuations in species dynamics. 

Amongst the many results that this promising technique grants access to, several findings from our investigation are worth recapping here. Part of the power of random matrix theory is that it uncovers universal properties of large classes of systems of a certain type. In the present case we find that, in analogy to the famous Wigner semi-circle law, the details of the distributions of matrix elements are unimportant beyond the handful of key parameters identified. Our parameter $\gamma$, which controls the proportion of predator-prey interactions (and hence the correlation of off-diagonal elements in the interaction matrix) is found to be of crucial importance. At one extreme, we find a semi-circular spectral profile, at the other we find a pole at zero frequency which has either $1/\omega^2$ or $1/\sqrt{\omega}$ divergence, depending on the symmetry of interactions. When an explicit trophic structure is incorporated into the model, it was necessary to adapt our method to general bipartite networks. Here, we found a gap in the power spectral density, implying that this high-level structure leads to greater long-term temporal stability. Finally, going beyond these universal results for the mean spectrum, we find a huge variability in fluctuations at the individual species level. These are not visible within the bulk but are captured by a single defect approximation, showing that some species may exhibit quasi-cyclic oscillations even when no such signal is present in the larger system. 

In each model investigated in this paper, we have characterised stochastic behaviour emerging on a macroscopic scale from the statistical properties of the underlying microscopic interactions. We emphasise that, as illustrated in our section on trophic levels, the fluctuations observed in large scale systems with a certain structure are likely to be substantially different from those of small scale models previously investigated. The models presented here have been chosen for simplicity and clarity, and they only scratch the surface of what can be achieved with this method. More realistic models might include a consideration of e.g. the dynamical assembling process of ecosystems~\cite{galla2018dynamically}, heterogeneous turnover rates~\cite{gibbs2018effect}, or explicitly spatial models where spatio-temporal patterning may persist \cite{mckane2014stochastic}. 

From an ecological perspective, it is desirable to connect our theoretical work to empirical investigations into ecosystem stability. In contrast to the traditional viewpoint of asymptotic linear stability, our methods directly address a fundamental empirical quantity -- timeseries of species abundance. Beyond simply providing more detail as to the temporal dynamics of an ecosystem around an equlibrium point, our method has also opened up the exciting possibility of identifying the signature of a certain interaction structures in the power spectrum of oscillations in data gathered in the field.
We fitted our model to a highly resolved time series data set on a plankton ecosystem. We found that the empirical data is indicative of an ecology dominated by competitive and mutalistic interactions, with far fewer predator-prey interactions. This insight in consistent with recent results that suggest that self‐regulation (competition) and facilitation (mutualism) are widespread in phytoplankton communities~\cite{picoche2020}.

In order to further realise this vision, some important further work is needed. Real ecosystems do not exist in a vacuum -- we must consider the role of the surrounding environment, including interactions with external factors such as seasonal variation or changing climate. Our theoretical approach encourages further work focused on the application to data sets gathered in field studies with modifications more suitable for the method we presented.

Finally, we wish to emphasise that ---despite the ecological focus in this paper--- the models of the kind we analysed are ubiquitous in many different fields, and the methods we use throughout the paper offer a general framework for large dynamical systems with random variables. Models of large interaction networks are also used in fields as varied as deep learning \cite{pennington2017nonlinear}, finance \cite{moran2019will}, biochemistry \cite{luo2007constructing} and neuroscience \cite{almog2019uncovering}. All of these systems depend on a high number of parameters that are often difficult to measure empirically. Our method provides a possibility to compute the power spectral density and gain insight into the model, which relies only on statistical meta parameters.


\section*{Methods}
\subsection*{Power Spectral Density for a General Ornstein-Uhlenbeck Process}

In the following we develop a method to compute the power spectral density of $N$-dimensional Ornstein-Uhlenbeck processes, 
\begin{equation}
    \frac{d\boldsymbol{\xi}}{dt} = \boldsymbol{A}\boldsymbol{\xi} + \boldsymbol{\zeta}(t),
    \label{eq_ornstein_uhlenbeck}
\end{equation}
where $\boldsymbol{\zeta}(t)$ is an $N$-vector of Gaussian white noise with correlations
$\mathbb{E}[\boldsymbol{\zeta}(t)\boldsymbol{\zeta}(t')^T] =\delta(t-t')\boldsymbol{B}$.
The matrix $\boldsymbol{A}$ determines the mean behaviour of $\boldsymbol{\xi}$ and is considered to be locally stable, i.e. all eigenvalues of $\boldsymbol{A}$ have negative real part.
Using the matrices $\boldsymbol{A}$ and $\boldsymbol{B}$ one can fully determine the power spectral density of fluctuations for the Ornstein-Uhlenbeck process.

We are interested in the case that the coefficients $A_{ij}$ and $B_{ij}$ are derived from a complex network of interactions with weights drawn at random, possibly with correlations. This framework encompasses a very general class of models with a wealth of real-world applications including but not limited to the ecological focus we have here.  The method we describe exploits the underlying network structure of $\bm{A}$ and $\bm{B}$ to deduce a self-consistent scheme of equations whose solution contains information on the power spectral density. 

We start with the definition of the power spectral density $\boldsymbol{\Phi}(\omega)$ as the Fourier transform of the covariance ${\mathbb{E}[ \boldsymbol{\xi}(t)\boldsymbol{\xi}(t+\tau)^T ]}$ at equilibrium,
\begin{equation}
    \boldsymbol{\Phi}(\omega) = \int_{-\infty}^{\infty} \mathrm{e}^{-\text{i}\omega\tau}\mathbb{E}[ \boldsymbol{\xi}(t)\boldsymbol{\xi}(t+\tau) ] d\tau.
    \label{eq_powerspectrum_definition_main}
\end{equation}
From \cite{gardiner_stochastic_2009} on multivariate Ornstein-Uhlenbeck processes, we know that the power spectral density can also be written in the form of the matrix equation,
\begin{equation}
    \boldsymbol{\Phi}(\omega) = (\boldsymbol{A}-i\omega\boldsymbol{I})^{-1} \boldsymbol{B} (\boldsymbol{A}^T+i\omega\boldsymbol{I})^{-1}.
    \label{eq_powerspectrum_definition}
\end{equation}
In practice, this equation is difficult to use for large systems as large matrix inversion is analytically intractable and numerical schemes are slow and sometimes unstable. We take an alternative route by recasting Eq.~(\ref{eq_powerspectrum_definition}) as a complex Gaussian integral reminiscent of problems appearing in the statistical physics of disordered systems. Our approach in the following is to treat $\omega$ as a fixed parameter and drop the explicit dependence from our notation. We begin by writing 
\begin{equation}
    \begin{split}
        \boldsymbol{\Phi}(\omega)=&\frac{|\boldsymbol{A}-i\omega\boldsymbol{I}|^2}{\pi^N|\boldsymbol{B}|} \int_{\mathbb{C}} e^{-\boldsymbol{u}^{\dagger}\boldsymbol{\Phi}^{-1}\boldsymbol{u}} \boldsymbol{u}\boldsymbol{u}^{\dagger} \prod_{i=1}^N du_i \,.
    \end{split}
\end{equation}
Simplification of the integrand is achieved by unpicking the matrix inversion in the exponent via a Hubbard-Stratonovich transformation~\cite{stratonovich1957, hubbard1959}. To this end we recast the system in the language of statistical mechanics by introducing $N$ complex-valued `spins' $u_i$ and $N$ auxiliary variables $v_i$, with the `Hamiltonian' 
\begin{equation}
    \mathcal{H}(\boldsymbol{u},\boldsymbol{v}) = -\boldsymbol{u}^\dagger(\boldsymbol{A}-\text{i}\omega)\boldsymbol{v} +\boldsymbol{v}^\dagger(\boldsymbol{A}-\text{i}\omega)^\dagger \boldsymbol{u} +\boldsymbol{v}^\dagger\boldsymbol{B}\boldsymbol{v} \,.
    \label{eq_hamiltonian}
\end{equation}
Introducing a bracket operator
\begin{equation}
    \langle \cdots\rangle := \frac{\int_{\mathbb{C}}e^{-\mathcal{H}(\boldsymbol{u},\boldsymbol{v})}(\cdots)d\boldsymbol{u}d\boldsymbol{v}}{\int_{\mathbb{C}}e^{-\mathcal{H}(\boldsymbol{u},\boldsymbol{v})}d\boldsymbol{u}d\boldsymbol{v}}\,,
\end{equation}
we can obtain succinct expressions for the power spectral density $\boldsymbol{\Phi}=\langle \boldsymbol{u}\boldsymbol{u}^\dag\rangle$ as well as the resolvent matrix $\boldsymbol{\mathcal{R}}=(\text{i}\omega-\boldsymbol{A})^{-1}=\langle \boldsymbol{u}\boldsymbol{v}^\dag\rangle$.
Thus we may write,
\begin{equation}
    \boldsymbol{\Phi}=\frac{1}{\mathcal{Z}} \int_{\mathbb{C}} e^{-\mathcal{H}(\boldsymbol{u},\boldsymbol{v})} \boldsymbol{u}\boldsymbol{u}^{\dagger} \prod_{i=1}^N du_idv_i \,,
    \label{eq_powerspec_hamiltonian}
\end{equation}
where ${\mathcal{Z}=|\boldsymbol{A}-i\omega\boldsymbol{I}|^2 / \pi^{2N}}$.

This construction may seem laborious at first, but it unlocks a powerful collection of statistical mechanics tools, including the {`cavity method'}.
Originally, the cavity method has been introduced in order to analyse a model for spin glass systems~\cite{mezard_sk_1986, mezard_spin_1987}. 
Further applications of the method include the analysis of the eigenvalue distribution in sparse matrices~\cite{rogers_cavity_2008, rogers_cavity_2009,metz2019spectral}. 
We will exploit the network structure in a similar fashion in order to compute the power spectral density.

In our analysis, we find that it is convenient to split the Hamiltonian in Eq.~(\ref{eq_hamiltonian}) into the sum of its local contributions at site $i$, $\mathcal{H}_i$,  and contributions from interactions between $i$ and $j$, $\mathcal{H}_{ij}$,
\begin{equation}
    \mathcal{H}=\sum_i\mathcal{H}_i +\sum_{i\sim j}\mathcal{H}_{ij} \,.
    \label{eq_hamiltonian_split}
\end{equation}
These terms can be decomposed as $\mathcal{H}_i=\boldsymbol{w}_i^\dagger\boldsymbol{\chi}_i\boldsymbol{w}_i$ and $\mathcal{H}_{ij}=\boldsymbol{w}_i^\dagger\boldsymbol{\chi}_{ij}\boldsymbol{w}_j$, where we introduce the compound spins $\boldsymbol{w}_i=(u_i,v_i)^T$ and transfer matrices,
\begin{equation}
    \begin{split}
        \boldsymbol{\chi}_i &= 
            \left(\begin{matrix}
                0 & A_{ii}+i\omega \\
                -A_{ii}+i\omega & B_{ii}
            \end{matrix}\right) \,,
        \\
        \boldsymbol{\chi}_{ij} &= 
            \left(\begin{matrix}
                0 & A_{ji} \\
                -A_{ij} & B_{ij}
            \end{matrix}\right) \,.
    \end{split}
    \label{eq_helping_matrices_def}
\end{equation}

Let us focus on the power spectral density of a particular variable $\xi_i$, obtained from the diagonal element $\phi_i=\Phi_{ii}$.
For this we compute the single-site marginal $f_i$ by integrating over all other variables,
\begin{equation}
    f_i(\boldsymbol{w}_i) = \frac{1}{\mathcal{Z}} \int_{\mathbb{C}}e^{-\mathcal{H}}\prod_{j\ne i} d\boldsymbol{w}_j.
    \label{eq_single_site_marginal}
\end{equation}
Alternatively, $\phi_i$ can be obtained as the top left entry of the covariance matrix $\boldsymbol{\Psi}_i=\langle \boldsymbol{w}_i\boldsymbol{w}_i^{\dagger} \rangle$.
We write the covariance matrix as the integral,
\begin{equation}
    \boldsymbol{\Psi}_i= \int_{\mathbb{C}} f_i(\boldsymbol{w}_i) \boldsymbol{w}_i\boldsymbol{w}_i^{\dagger} d\boldsymbol{w}_i \,,
    \label{eq_covariance_matrix_a}
\end{equation}
which could also be expressed in terms of a Gaussian integral,
\begin{equation}
    \boldsymbol{\Psi}_i= \frac{1}{\pi^2|\boldsymbol{\Psi}_i|} \int_{\mathbb{C}} e^{-\boldsymbol{w}_i^{\dagger}\boldsymbol{\Psi}_i^{-1}\boldsymbol{w}_i} \boldsymbol{w}_i\boldsymbol{w}_i^{\dagger} d\boldsymbol{w}_i \,.
    \label{eq_covariance_matrix_b}
\end{equation}
By comparing Eqs.~(\ref{eq_covariance_matrix_a}) and (\ref{eq_covariance_matrix_b}) we find that
\begin{equation}
    f_i(\boldsymbol{w}_i)=\frac{1}{\pi^2|\boldsymbol{\Psi}_i|} e^{-\boldsymbol{w}_i^{\dagger}\boldsymbol{\Psi}_i^{-1}\boldsymbol{w}_i} \,.
    \label{eq_single_site_marginal_covariance_matrix}
\end{equation}
We now insert Eq.~(\ref{eq_hamiltonian_split}) into Eq.~(\ref{eq_single_site_marginal}) and obtain,
\begin{equation}
    f_i(\boldsymbol{w}_i) = \frac{1}{\pi^2|\boldsymbol{\Psi}_i|} e^{-\mathcal{H}_i} \int_{\mathbb{C}} \prod_{i\sim j} \left( e^{-\mathcal{H}_{ij}-\mathcal{H}_{ji}} f_j^{(i)} d\boldsymbol{w}_j \right) \,,
    \label{eq_single_site_marginal_recursion}
\end{equation}
where we write $f_j^{(i)}$ for the {`cavity marginals'},
\begin{equation}
    f_j^{(i)}(\boldsymbol{w}_j) = \frac{1}{\mathcal{Z}^{(i)}} \int_{\mathbb{C}} e^{-\mathcal{H}^{(i)}}\prod_{k\ne i,j} d\boldsymbol{w}_k \,.
    \label{eq_single_site_marginal_cavity}
\end{equation}
In essence, the above discussion amounts to organising the $2N$ integrals in Eq.~(\ref{eq_powerspec_hamiltonian}) in a convenient way, with the advantage of providing a simple intuition for the role of the underlying network. The superscript $(i)$ is used to indicate that the quantity corresponds to the cavity network where node $i$ has been removed. We will further use this notation for the {`cavity covariance matrix'} $\boldsymbol{\Psi}_{jl}^{(i)}$ introduced in the following.

Next we perform the integration in Eq.~(\ref{eq_single_site_marginal_recursion}) and compare to the form in Eq.~(\ref{eq_single_site_marginal_covariance_matrix}).
We thus obtain a recursion formula for the covariance matrix $\boldsymbol{\Psi}_i$ and the cavity covariance matrices $\boldsymbol{\Psi}_{jl}^{(i)}$,
\begin{equation}
    \boldsymbol{\Psi}_i = \left( \boldsymbol{\chi}_i - \sum_{\substack{i\sim j \\ i\sim l}}\boldsymbol{\chi}_{ij} \boldsymbol{\Psi}_{jl}^{(i)} \boldsymbol{\chi}_{li} \right)^{-1} \,,
    \label{eq_psi_general}
\end{equation}
where the notation $i\sim j$ indicates that we sum over nodes $j$ connected to node $i$. 
Unless there is some specific structure underlying the network, we assume that most real world cases have a {`tree-like'} structure from the local view point of a single node $i$.
Hence, it is highly unlikely that the nodes $j$ and $l$ are nearby in the cavity network where node $i$ is removed, and thus $\boldsymbol{\Psi}_{jl}^{(i)}$ only gives non-zero contributions if $j=l$.
We therefore reduce Eq.~(\ref{eq_psi_general}) and obtain for the covariance matrix,
\begin{equation}
    \boldsymbol{\Psi}_i = \left( \boldsymbol{\chi}_i - \sum_{i\sim j}\boldsymbol{\chi}_{ij} \boldsymbol{\Psi}_{j}^{(i)} \boldsymbol{\chi}_{ji} \right)^{-1}.
    \label{eq_psi_cavity_a}
\end{equation}
Similarly, the cavity covariance matrix obeys the equation,
\begin{equation}
    \boldsymbol{\Psi}_j^{(i)} = \left( \boldsymbol{\chi}_j - \sum_{j\sim k, k\ne i}\boldsymbol{\chi}_{jk} \boldsymbol{\Psi}_{k}^{(j)} \boldsymbol{\chi}_{kj} \right)^{-1}.
    \label{eq_psi_cavity_b}
\end{equation}
Here we use that $\boldsymbol{\Psi}^{(i,j)}=\boldsymbol{\Psi}^{(j)}$ when the nodes $i$ and $k$ are not connected.
In other words, removing node $j$ from the cavity network where node $i$ is missing, has the same effect as removing it from the full network. The system in Eq.~(\ref{eq_psi_cavity_a}) describes a collection of nonlinear matrix equations that must be solved self-consistently. 

For networks with high enough connectivity (and to good approximation even with modest connectivity), the removal of a single node does not affect the rest of the network, as its contribution is negligible compared to the full system. Hence the system in Eq.~(\ref{eq_psi_cavity_a}) can be reduced to a smaller set of equations approximately satisfied by the matrices $\bm{\Psi}_i$:
\begin{equation}
    \boldsymbol{\Psi}_i \approx \left( \boldsymbol{\chi}_i - \sum_{i\sim j}\boldsymbol{\chi}_{ij} \boldsymbol{\Psi}_{j} \boldsymbol{\chi}_{ji} \right)^{-1}.
    \label{eq_psi_tap}
\end{equation}
The power spectral density $\phi_i$ can be obtained as the top left entry of $\boldsymbol{\Psi}_i$.

In order to progress further, we now consider specific approximations that help us compute the power spectral density.
First we take a mean-field approach in order to obtain the mean power spectral density for all nodes part of the network; we then use the result for the mean field in order to compute a close approximation to the local power spectral density of a single node.
Later, we adapt the method to partitioned networks where nodes belong to different types of connected groups.

\paragraph{Mean Field}
For the following we assume that all agents in the system behave the same on average.
In practice, the terms governed by self-interactions $A_{ii}$ are drawn from the same distribution for all agents.
Similarly, the terms including $B_{ii}$ are governed by one distribution.
Interaction strengths and connections with other nodes in the network are also sampled equally for all agents (we have explored a large Lotka-Volterra ecosystem as an example of such a network).
In the mean-field (MF) formulation we assume that the mean degree and excess degree are approximately equal, and replace all quantities in Eqs.~(\ref{eq_psi_cavity_a}) and (\ref{eq_psi_cavity_b}) with their average.  $\boldsymbol{\Psi}_i=\boldsymbol{\Psi}^{\mathrm{MF}} \,\forall i$.
We then obtain the following recursion equation,
\begin{equation}
    \boldsymbol{\Psi}^{\mathrm{MF}} = \left[ \mathbb{E}[\boldsymbol{\chi}_i] - \mathbb{E}\left( \sum_{i\sim j} \boldsymbol{\chi}_{ij}\boldsymbol{\Psi}^{\mathrm{MF}} \boldsymbol{\chi}_{ji} \right)\right]^{-1}.
    \label{eq_psi_recursion_meanfield}
\end{equation}
In order to solve this equation, we parameterise, 
\begin{equation}
    \boldsymbol{\Psi}^{\mathrm{MF}} =
        \left(\begin{matrix}
            \phi & r \\
            -\bar{r} & 0
        \end{matrix}\right),
    \label{eq_parameterisation_meanfield}
\end{equation}
where the top left entry $\phi$ corresponds to the mean power spectral density, and we introduce $r$ as the mean diagonal element of the resolvent matrix $\bm{\mathcal{R}}$. 
Finally by inserting the ansatz of Eq.~(\ref{eq_parameterisation_meanfield}) into Eq.~(\ref{eq_psi_recursion_meanfield}) we obtain,
\begin{equation}
    \begin{split}
        &\left(\begin{matrix}
            \phi & r \\
            -\bar{r} & 0
        \end{matrix}\right) ^{-1}
        =
        \left(\begin{matrix}
            0 & \mathbb{E}[A_{ii}]+i\omega \\
            -\mathbb{E}[A_{ii}]+i\omega & \mathbb{E}[B_{ii}]
        \end{matrix}\right) 
        \\ &+c
        \left(\begin{matrix}
            0 & \bar{r} \mathbb{E}[A_{ij}A_{ji}] \\
            -r \mathbb{E}[A_{ij}A_{ji}] & \phi \mathbb{E}[A_{ij}^2] + (r+\bar{r}) \mathbb{E}[A_{ij}B_{ij}]
        \end{matrix}\right),
    \end{split}
    \label{eq_sr_matrix_highconn}
\end{equation}
where $c$ is the average degree (i.e. number of connections) per node.
Moreover, the expectations in the second term are to be taken over connected nodes $i\sim j$ (i.e. non-zero matrix entries).

From Eq. (\ref{eq_sr_matrix_highconn}) above, we obtain the equations, 
\begin{equation}
    \begin{split}
        \frac{\phi}{|r|^2} &= \mathbb{E}[B_{ii}] +c \left(\phi\mathbb{E}[A_{ij}^2] + 2\mathrm{Re}(r) \mathbb{E}[A_{ij}B_{ij}] \right), \\
        \frac{\bar{r}}{|r|^2} &= -\mathbb{E}[A_{ii}] +i\omega -cr\mathbb{E}[A_{ij}A_{ji}].
    \end{split}
    \label{eq_sr_highconn}
\end{equation}
We solve the second equation in Eq.~(\ref{eq_sr_highconn}) for $r$ and write the mean power spectral density in terms of $r$,
\begin{equation}
    \begin{split}
        \phi =& |r|^2\frac{\mathbb{E}[B_{ii}] +2c\mathrm{Re}(r)\mathbb{E}[A_{ij}B_{ij}]}{1-c|r|^2\mathbb{E}[A_{ij}^2]}, \\
        r =& \frac{1}{2c\mathbb{E}[A_{ij}A_{ji}]} \left[-\mathbb{E}[A_{ii}] +i\omega \right. \\
            &\left. - \sqrt{(-\mathbb{E}[A_{ii}]+i\omega)^2 -4c\mathbb{E}[A_{ij}A_{ji}]}\right]
    \end{split}
    \label{eq_sr_meanfield}
\end{equation}

This equation informs the first part of the results presented in the main text. 

\paragraph{Single Defect Approximation}
The Single Defect Approximation (SDA) makes use of the mean-field approximation for the cavity fields, but retains local information about individual nodes. 
We parameterise similarly to Eq.~(\ref{eq_parameterisation_meanfield}) for a single individual. 
Moreover, we replace all other quantities with the respective mean-field approximation.
Specifically, we obtain
\begin{equation}
    \begin{split}
        &\left(\begin{matrix}
            \phi_i^{\text{SDA}} & r_i^{\text{SDA}} \\
            -\bar{r}_i^{\text{SDA}} & 0
        \end{matrix}\right)^{-1} 
        = \left(\begin{matrix}
            0 & A_{ii}+i\omega \\
            -A_{ii}+i\omega & B_{ii}
        \end{matrix}\right) \\
        &+\sum_{i\sim j}
        \left(\begin{matrix}
            0 & \bar{r}^{\text{MF}} A_{ij}A_{ji} \\
            -r^{\text{MF}} A_{ij}A_{ji} & \phi^{\text{MF}} A_{ij}^2 +(r^{\text{MF}}+\bar{r}^{\text{MF}})A_{ij}B_{ij} 
        \end{matrix}\right) \,.
    \end{split}
\end{equation}
We solve this equation for $\phi_i^{\text{SDA}}, r_i^{\text{SDA}}$, which delivers
\begin{equation}
    \begin{split}
        \frac{\phi_i^{\text{SDA}}}{|r_i^{\text{SDA}}|^2} &=     
             \phi^{\text{MF}}\sum_{i\sim j}A_{ij}^2 + 2\mathrm{Re}(r^{\text{MF}})\sum_{i\sim j}A_{ij}B_{ij}+ B_{ii}  \,,\\
        r_i^{\text{SDA}} &= \left( A_{ii}+i\omega+\bar{r}^{\text{MF}}\sum_{i\sim j}A_{ij}A_{ji} \right)^{-1} \,.
    \end{split}
\end{equation}

\paragraph{Partitioned Network}
Previously we assumed that all nodes in a network are interchangeable in distribution. 
However, many real-world applications feature agents with different properties, imposing a high-level structure on the network. 
We realise this by partitioning nodes into distinct groups that interact with each other (see the section {Trophic Structure Model} for a simple example).

In order to handle different connected groups we make use of the cavity method as in Eqs.~(\ref{eq_psi_cavity_a}) and (\ref{eq_psi_cavity_b}).
In particular, we split the sum in the second term on the right-hand side of these equations into contributions from each group in the partitioned network.
Let $M$ denote the number of subgroups $V_m$ in a partitioned network then we write,
\begin{equation}
    \begin{split}
        \boldsymbol{\Psi}_i &= \left(\boldsymbol{\chi}_i -\sum_m^M\sum_{\substack{i\sim j \\ j\in V_m}} \boldsymbol{\chi}_{ij}\boldsymbol{\Psi}_j^{(i)}\boldsymbol{\chi}_{ji}\right)^{-1} \,, \\
        \boldsymbol{\Psi}_j^{(i)} &= \left(\boldsymbol{\chi}_j -\sum_m^M\sum_{\substack{j\sim k \\ k\in V_m}} \boldsymbol{\chi}_{jk}\boldsymbol{\Psi}_k^{(j)}\boldsymbol{\chi}_{kj}\right)^{-1} \,.
    \end{split}
    \label{eq_powerspec_recursion_partition}
\end{equation}
Similar to the previous sections 
we replace all quantities with a mean-field average $\boldsymbol{\Psi}_m^{\mathrm{MF}}$, but for each group separately.
Hence we obtain $M$ equations of the form
\begin{equation}
    \boldsymbol{\Psi}_i^{\mathrm{MF}} = \left[ \mathbb{E}[\boldsymbol{\chi}_i] - \mathbb{E}\left( \sum_m^M\sum_{\substack{i\sim j \\ j\in V_m}} \boldsymbol{\chi}_{ij}\boldsymbol{\Psi}_m^{\mathrm{MF}} \boldsymbol{\chi}_{ji} \right)\right]^{-1}.
    \label{eq_psi_partition_meanfield}
\end{equation}
In order to compute the mean power spectral density for different groups separately, we use a parameterisation as in Eq.~(\ref{eq_parameterisation_meanfield}) for each group.
Therefore we have,
\begin{equation}
    \boldsymbol{\Psi}_m^{\mathrm{MF}} =
        \left(\begin{matrix}
            \phi_m & r_m \\
            -\bar{r}_m & 0
        \end{matrix}\right),
\end{equation}
for all $m=1,\dots,M$.
This delivers $2M$ equations to solve for all $r_m$ and $\phi_m$.
Numerically this is straight forward, although algebraically long-winded for the general case.
However, the equations simplify for special cases.
In the section {Trophic Structure Model} we demonstrate this method for a bipartite network where a lack of intra-group interactions simplifies the analysis.

\subsection*{Large Lotka-Volterra Ecosystem}

\paragraph{Model Description}

First, we define the framework for a general Lotka-Volterra ecosystem with $N$ species and a large but finite system size $V\gg1$. Note that this parameter can be interpreted as a scaling factor for the fluctuation amplitude and thus, larger systems exhibit higher stability and quantitative reliability for our analytic results.
Let $X_i$ denote the number of individuals and $x_i=X_i/V$ the density of species $i=1,\dots,N$.
We start from the following set of reactions that define the underlying stochastic dynamics of the system:
\begin{equation}
\begin{split}
&X_i \overset{b_i}{\longrightarrow} 2X_i \quad \text{(birth)}\\
& 2X_i \overset{R_{ii}}{\longrightarrow} X_i \quad \text{(death)}\\
&X_i + X_j \overset{R_{ij}}{\longrightarrow}
         \begin{cases}
         2X_i+X_j &\mathrm{(mutualism)},\\
         X_i &\mathrm{(competition)},\\
         2X_i &\mathrm{(predation)}.
         \end{cases}
\end{split}
\end{equation}

The self-interactions are governed by the birth rate ${b_i>0}$ and density-dependent mortality rate ${R_{ii}>0}$.
Furthermore, we define three interaction types between species $i$ and $j$, namely mutualism, competition and predation. 
In the case of mutualistic interactions, both species benefit from each other, whereas competition means that both species have a higher mortality rate, depending on the density of the other species.
For predator-prey pairs, one predator species benefits from the death of a prey species.
The predator and prey species are chosen randomly, such that species $i$ is equally likely to be a predator or prey of species $j$.

With probability $P_c$ we assign an interaction rate $R_{ij}>0$ to the species pair $(i,j)$, and with probability $1-P_c$ there is no interaction between species $i$ and $j$ (i.e. $R_{ij}=0$). 
In other words, each species has on average $c=NP_c$ interaction partners.
The reaction rates are considered to be i.i.d. random variables drawn from a half-normal distribution $|\mathcal{N}(0,\sigma^2)|$, where we write for the mean reaction rate $\mu=\mathbb{E}[R_{ij}]=\sigma\sqrt{2/\pi}$ and raw second moment $\sigma^2=\mathbb{E}[R_{ij}^2]$. 
For each interaction pair, the interaction type is chosen such that the proportion of predator-prey pairs is $p\in[0,1]$, and all non-predator-prey interactions are equally distributed between mutualistic and competitive interactions (i.e. the overall proportion of mutualistic/competitive interactions is $1/2(1-p)$).
Lastly, we define the symmetry parameter $\gamma=1-2p$, where $\gamma=-1$ if all interactions are of predator-prey type ($p=1$), and similarly $\gamma=+1$ if there are no predator-prey interactions ($p=0$).
In a mixed case where predator-prey and mutualistic/competitive interactions have equal proportion ($p=1/2$), we have $\gamma=0$.
Later we will see that $\gamma$ is equivalent to the correlation of signed interaction strengths.

In the limit $V\rightarrow\infty$, the dynamics of the species density $x_i$ obey the ordinary differential equations,
\begin{equation}
    \frac{dx_i}{dt} = x_i \left( b_i + \sum_{j}^N \alpha_{ij}x_j \right),
    \label{eq_deterministic_LV}
\end{equation}
where $\alpha_{ij}$ are the interaction coefficients with ${|\alpha_{ij}|=|\alpha_{ji}|=R_{ij}}$.
The signs of the interaction coefficients are determined by the type of interaction between species $i$ and $j$.
For mutualistic interactions we have ${\alpha_{ij}=\alpha_{ji}>0}$, and ${\alpha_{ij}=\alpha_{ji}<0}$ for competitive interactions.
In the case of predator-prey interactions the coefficients have opposite sign ${\alpha_{ij}=-\alpha_{ji}}$.
Hence the symmetry parameter as described above is given by the correlation of interaction coefficients $\gamma=\mathbb{E}[\alpha_{ij}\alpha_{ji}]$.
Furthermore, in order to ensure bounded species densities, we require negative self-interactions $\alpha_{ii}=-R_{ii}<0$. 

If species live in isolation (i.e. when ${\alpha_{ij}=0 \,\forall i\ne j}$), we see that the densities approach the {`effective'} carrying capacity $K_i=-b_i/\alpha_{ii}$.
For the following computations we consider a large Lotka-Volterra system. 
Since we are only interested in the effects of interactions between species, we assume that all self-interactions are approximately equal.
Thus we write for the birth rate $b_i=b$ and mortality rate $\alpha_{ii}=-b$.
This gives the effective carrying capacity $K=1$ for all species.

The fixed point $\boldsymbol{x}^*$ at the deterministic equilibrium state is given by,
\begin{equation}
    x_i^* = 1 + \sum_{j \ne i}\alpha_{ij}x_j^*.
\end{equation}
We assume a random mixture of mutualistic and competitive interactions with equal proportions, and therefore the interaction coefficients $\alpha_{ij}$ have zero mean ($\forall i\ne j$).
Furthermore, we postulate that for large ecosystems where $N\rightarrow\infty$, the equilibrium state $x_i^*=\mathbb{E}[x_i^*]\equiv x^*$.
Hence we obtain the expected equilibrium density $x^* = 1$ for all species $i$. Note that the following computations are valid for any known fixed point $\boldsymbol{x}^*$, and our assumptions are for mathematical simplification only. The results are independent of the particular equilibrium configuration, as long as a stable equilibrium can be measured and extracted from data (we discuss a few caveats where we apply our method to time series data from a plankton ecosystem).
This assumption allows us to write the Jacobian matrix for a linearisation around the equilibrium state, with elements,
\begin{equation}
    \begin{split}
    J_{ii}|_{\boldsymbol{x}=\boldsymbol{x}^*} &= \alpha_{ii} = -b, \\
    J_{ij}|_{\boldsymbol{x}=\boldsymbol{x}^*} &= \alpha_{ij}.
    \end{split}
\end{equation}
In other words, the community matrix of a large Lotka-Volterra system as described above has the same form as the interaction matrix, i.e. $A_{ij}=\alpha_{ij}$.
The local stability of such community matrix $A$ is given by the elliptic law~\cite{allesina_stability_2012, allesina_stabilitycomplexity_2015}.
It states that with high probability all eigenvalues of the random matrix $A$ are distributed on an ellipse in the complex plane, centered at $(-b,0)$ on the real axis.
Thus for a stable matrix we require all eigenvalues to be negative, and hence the horizontal semi-axis of the ellipse determines the allowed range for the centre.
It follows the stability criterion,
\begin{equation}
    \sqrt{c\sigma^2} (1 +\gamma) <b,
    \label{eq_stability_criteria}
\end{equation}
with the average number of connections $c$ per species, and the correlation $\gamma=\mathbb{E}[A_{ij}A_{ji}]$.
For a random community matrix (i.e. $\gamma=0$), we recover the stability criterion that has been proven by May~\cite{may_will_1972}, $\sqrt{c\sigma^2}<b$.
If $\gamma<0$, where the proportion of predator-prey type interactions is larger, the horizontal semi-axis of the ellipse becomes smaller.
In other words, the stability criterion relaxes for predator-prey interactions.
For $\gamma=-1$ (i.e. $A_{ij}=-A_{ji} \forall i,j$), all interactions are of predator-prey type and all eigenvalues become purely imaginary.
Therefore the stability criterion becomes $0<b$, as the ellipse stretches vertically into the imaginary plane.
The opposite is true for mutualistic/competitive interactions (i.e. $\gamma=+1$), where eigenvalues are distributed on an ellipse with large horizontal radius along the real axis.
Thus it is more likely that some eigenvalues have positive real part and the system destabilises.
We choose the parameter $b$ for each case, such that the stability criteria are fulfilled.

For a large but finite system size $V$, we write the stochastic differential equations,
\begin{equation}
    \frac{dx_i}{dt} = x_i \left( b +\sum_{j}^N \alpha_{ij}x_j \right) + \frac{1}{\sqrt{V}} \eta_i(t),
    \label{eq_HLV_sde}
\end{equation}
where $\eta_i(t)$ are Gaussian random variables with $\langle\eta_i(t),\eta_j(t')\rangle=\delta(t-t')B_{ij}$.
The noise matrix $\boldsymbol{B}$ can be obtained from the reactions that determine the process.
The diagonal elements are given by the self-interactions and total interaction from all other species, and the off-diagonal elements depend on the type of interaction between species $i$ and $j$. We assume that only predator-prey type interactions contribute to the covariance of species fluctuations (i.e. that only predator-prey interactions involve the simultaneous change in abundance of a species pair). 
Therefore, we write
\begin{equation}
    \begin{split}
        B_{ii}(\bm{x}) &= x_i \left( b + \sum_{j=1}^N R_{ij}x_j \right),
        \\
        B_{ij}(\bm{x}) &= \begin{cases}
            -R_{ij}x_i x_j  &\text{if }\alpha_{ij}=-\alpha_{ji}, \\
            0 &\text{else.}
            \end{cases}
    \end{split}
    \label{eq_HLV_B1}
\end{equation}

We next linearise around the fixed point to obtain a new equation for the fluctuations, $\boldsymbol{\xi}=\sqrt{V}(\boldsymbol{x}-\boldsymbol{x}^*)$, which has the form of an Ornstein-Uhlenbeck process as defined in Eq.~(\ref{eq_ornstein_uhlenbeck}). Recall that in our simplified model the equilibrium abundance $\bm{x}^*=\bm{1}$ (note however, that in general the entries of the noise matrix $\boldsymbol{B}$ depend on the particular fixed point of a given system). Therefore we write for the noise matrix evaluated at the fixed point
\begin{equation}
    \begin{split}
        B_{ii}(\bm{x}^*) & = 2b +c\mu,
        \\
        B_{ij}(\bm{x}^*) &= \begin{cases}
              -R_{ij} &\text{if }\alpha_{ij}=-\alpha_{ji}, \\
            0 &\text{else,}
            \end{cases}
    \end{split}
    \label{eq_HLV_B}
\end{equation}
where $\mu$ is given as the mean reaction rate $\mu=\mathbb{E}[R_{ij}]=\sigma\sqrt{2/\pi}$.

\paragraph{Computing the Power Spectral Density}
Let us now compute the mean power spectral density $\phi$ of the process described above 
using Eq.~(\ref{eq_sr_meanfield}) as starting point.
We replace the necessary quantities that we obtain from the community matrix $\boldsymbol{A}$ and noise matrix $\boldsymbol{B}$ as defined in the previous section.
In particular, we have the expected diagonal elements of the community matrix $\mathbb{E}[A_{ii}]=-b$, and the noise matrix $\mathbb{E}[B_{ii}]=2b+c\mu$.
Moreover the raw second moment of the non-zero interactions is given by $\mathbb{E}[A_{ij}]=\sigma^2$ and the correlation $\mathbb{E}[A_{ij}A_{ji}]=\gamma\sigma^2$.
We use that ${\mathbb{E}[A_{ij}B_{ij}]=0 \,\forall i,j}$ since the off-diagonal elements of the noise matrix are only non-zero if there is a predator-prey interaction between species $i$ and $j$. 
However, the elements of $A_{ij}$ have opposite signs in the case of predator-prey pairs and thus sum to zero.

Plugging in these quantities into Eq.~(\ref{eq_sr_meanfield}) we obtain,
\begin{equation}
    \begin{split}
        \phi =& |r|^2\frac{2b+c\mu}{1-|r|^2c\sigma^2} \,, \\
        r =& \frac{1}{2c\gamma\sigma^2} \left[b +i\omega - \sqrt{(b+i\omega)^2 -4c\gamma\sigma^2}\right] \,.
    \end{split}
    \label{eq_HLV_powerspec}
\end{equation}
In the main text we explore the theoretical ecological consequences of this result. 

\subsection*{Trophic Structure Model}
\label{sec_a_BPP}

\begin{figure}[t]
    \centering
    \includegraphics[width=0.4\textwidth]{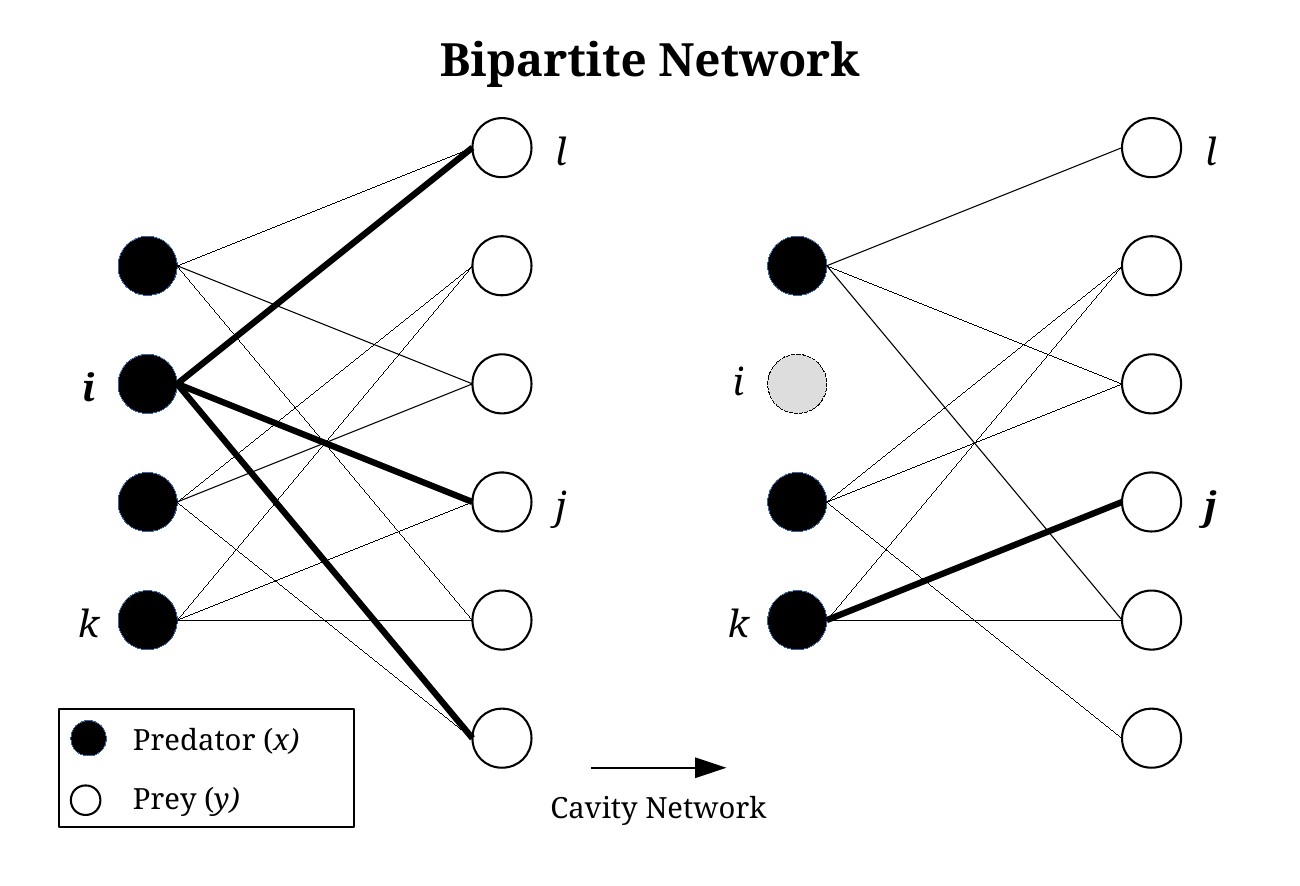}
    \caption{Example of a bipartite predator-prey network.
    Here we have $n_x=4$ predators with $c_x=3$ prey each, and $n_y=6$ prey with $c_y=2$ predators each.
    The bold lines illustrate the connections to the focal node before and after removing a node from the network.
    Predator nodes (black) only have local contributions from prey nodes (white) and vice versa.
    In the mean-field approximation for the power spectral density, these contributions are replaced by the average of each group.}
    \label{fig_bipartite_network}
\end{figure}

\paragraph{Model Description}
In the following we define a model analogous to the one described in the previous section. 
For a large but finite system size $V$ we write the model in terms of a stochastic process.
Previously we allowed for different types of interactions, however in this model we only focus on predator-prey interactions.
More specifically, the interaction network is partitioned into $N_x$ predator species and $N_y$ prey species, where $N=N_x+N_y$ is the total number of species.
We assume that predators only interact with prey and vice versa (i.e. we assume no inter-species interactions within the groups of predators or prey) as illustrated in Fig. \ref{fig_bipartite_network}. 
Moreover, each predator and prey species interacts with themselves (density-dependent mortality).

In the previous model we assigned the same birth rate to all species in the ecosystem.
Here we assume that predators decline at rate $d$ in absence of prey, and $b$ is the birth rate of prey species.
For simplicity, we assume that $d,b$ are fixed quantities, equal for all predators and prey respectively.
Furthermore $R_{ij}$ is the interaction rate between predator $i$ and prey $j$.
Each predator species has a fixed number of prey $c_x$ and each prey species has a fixed number of predators $c_y$, such that $N_xc_x=N_yc_y$.
The parameters $c_x,c_y$ can be interpreted as outgoing degrees of predator and prey nodes respectively.
Connections between predators and prey are then wired randomly.
The interaction strength is set to $R_{ij}=\alpha$, and  considered equal for all predator-prey interactions (analogous to a mean reaction rate).
Where there is no interaction between species, the interaction rate is simply set to zero.
Note that this means that the total sum of interaction strength is constant $\alpha c_x$ and $\alpha c_y$ for all predator and prey species respectively.
In contrast to the previous model, now only the network structure contributes to the randomness of the system.

Let $x_i$ denote the density of predator species ${i=1,\dots,N_x}$, and $y_j$ the density of prey species ${j=1,\dots,N_y}$.
In the deterministic limit where $V\rightarrow\infty$ we then write the following ODEs,
\begin{equation}
    \begin{split}
        \frac{dx_i}{dt} &= x_i \left(-d -x_i + \sum_{j=1}^{N_y} R_{ij}y_j \right), \\
        \frac{dy_j}{dt} &= y_j \left(b -y_j - \sum_{i=1}^{N_x} R_{ji}x_i \right). 
    \end{split}
    \label{eq_bipartite_ode}
\end{equation}
Given the fixed number of connections $c_x,c_y$ and interaction strength $\alpha$, we can simplify the ODEs to two equations for the average predator and prey densities,
\begin{equation}
    \begin{split}
        \frac{dx}{dt} &= x \left(-d -x + c_x\alpha y \right), \\
        \frac{dy}{dt} &= y \left(b -y - c_y\alpha x \right).
    \end{split}
    \label{eq_bipartite_ode_twoD}
\end{equation}
In the limit of large $N$, the equilibrium state of the system converges to the average quantities obtained from this reduced form.
The biologically relevant equilibrium states for this system are given by the trivial fixed points ${(x^*,y^*)=(0,0),(0,b)}$, and the non-trivial fixed point,
\begin{equation}
    \begin{split}
        x^* = \frac{c_x\alpha b-d}{c_xc_y\alpha^2+1}, \\
        y^* = \frac{c_y\alpha d+b}{c_xc_y\alpha^2+1}.
    \end{split}
    \label{eq_BPP_equilibrium}
\end{equation}

Next, we write the Jacobian matrix for a linearisation around the non-trivial fixed point.
The community matrix takes the form,
\begin{equation}
    \boldsymbol{A} = \left(\begin{matrix}
        -x^*\boldsymbol{I} & x^*\boldsymbol{R} \\
        -y^*\boldsymbol{R}^T & -y^*\boldsymbol{I}
    \end{matrix}\right),
\end{equation} 
where the first ${i=1,\dots,N_x}$ rows and columns represent the predator species, and the remaining ${j=N_x+1,\dots,N_x+N_y}$ rows and columns correspond to the prey species.

For a large but finite system size $V$ we write the corresponding stochastic differential equations,
\begin{equation}
    \begin{split}
        \frac{dx_i}{dt} &= x_i \left(-d -x_i + \sum_j^{N_y} R_{ij}y_j \right) +\frac{1}{\sqrt{V}}\eta_i(t), \\
        \frac{dy_j}{dt} &= y_j \left(b -y_j - \sum_i^{N_x} R_{ji}x_i \right) +\frac{1}{\sqrt{V}}\eta_j(t), \\
    \end{split}
    \label{eq_bipartite_sde}
\end{equation}
where $\eta_{i,j}(t)$ are Gaussian noise with $\langle\eta_i(t),\eta_j(t')\rangle=\delta_{ij}(t-t')B_{ij}$.
The noise matrix is given by the self- and total interactions on the diagonal, and the interactions between predators and prey on the off-diagonal.
We therefore write,
\begin{equation}
    \boldsymbol{B} = \left(\begin{matrix}
        2x^*(x^*+d)\boldsymbol{I} & -x^*y^*\boldsymbol{R} \\
        -x^*y^*\boldsymbol{R}^T & 2y^*b\boldsymbol{I}
    \end{matrix}\right).
\end{equation} 
Again, this allows us to write the dynamics in form of an Ornstein-Uhlenbeck process as defined in Eq.~(\ref{eq_ornstein_uhlenbeck}).

\paragraph{Computing the Power Spectral Density}
In the following, we use features of the bipartite interaction network.
For instance, all nodes that are connected to e.g. node $x_i$, will be prey nodes $y_j$, and thus are not connected with each other (see Fig. \ref{fig_bipartite_network}).
This allows us to write the following recursion formulas for the mean power spectral densities according to Eq.~(\ref{eq_psi_partition_meanfield}), 
\begin{equation}
    \begin{split}
        \boldsymbol{\Psi}_x^{-1} &= \mathbb{E}[\boldsymbol{\chi}_i] -\mathbb{E}\left[\sum_{i\sim j}^{N_y}\boldsymbol{\chi}_{ij}\boldsymbol{\Psi}_y\boldsymbol{\chi}_{ji} \right] \,, \\
        \boldsymbol{\Psi}_y^{-1} &= \mathbb{E}[\boldsymbol{\chi}_j] -\mathbb{E}\left[\sum_{j\sim i}^{N_x}\boldsymbol{\chi}_{ji}\boldsymbol{\Psi}_x\boldsymbol{\chi}_{ij} \right] \,.
    \end{split}
    \label{eq_powerspec_recursion_BPP_compact}
\end{equation}
Recall that the top left entries of $\boldsymbol{\Psi}_x$ and $\boldsymbol{\Psi}_y$ deliver the mean power spectral densities for predators $\phi_x$ and prey $\phi_y$ respectively. 
For the bipartite model, the helping matrices $\boldsymbol{\chi}_i, \boldsymbol{\chi}_{ij}$ (as defined in Eq.~(\ref{eq_helping_matrices_def})) are given by,
\begin{equation}
    \begin{split}
        \boldsymbol{\chi}_x &= \left(\begin{matrix}
            0 & -x+i\omega \\
            x+i\omega & 2x(x+d)
        \end{matrix}\right),
        \\
        \boldsymbol{\chi}_y &= \left(\begin{matrix}
            0 & -y+i\omega \\
            y+i\omega & 2yb
        \end{matrix}\right),
        \\
        \boldsymbol{\chi}_{xy} &= \left(\begin{matrix}
            0 & -\alpha y \\
            -\alpha x & -\alpha xy
        \end{matrix}\right),
        \quad
        \boldsymbol{\chi}_{yx} = \left(\begin{matrix}
            0 & \alpha x \\
            \alpha y & -\alpha xy
        \end{matrix}\right).
    \end{split}
\end{equation}
Inserting and writing out Eq.~(\ref{eq_powerspec_recursion_BPP_compact}) gives,
\begin{equation}
    \begin{split}
        \left(\begin{matrix}
            \phi_{x} & r_{x} \\
            -\bar{r}_x & 0
        \end{matrix}\right)^{-1} 
        =& \left(\begin{matrix}
            0 & -x+i\omega \\
            x+i\omega & 2x(x+d)
        \end{matrix}\right) \\
        &+\alpha^2c_x
        \left(\begin{matrix}
            0 & -\bar{r}_y xy \\
            r_y xy & \phi_y x^2 -(r_y+\bar{r}_y)x^2y
        \end{matrix}\right) \,, \\
        \left(\begin{matrix}
            \phi_{y} & r_{y} \\
            -\bar{r}_y & 0
        \end{matrix}\right)^{-1} 
        =& \left(\begin{matrix}
            0 & -y+i\omega \\
            y+i\omega & 2yb
        \end{matrix}\right) \\
        &+\alpha^2 c_y
        \left(\begin{matrix}
            0 & -\bar{r}_x xy \\
            r_x xy & \phi_x y^2 +(r_x+\bar{r}_x)xy^2
        \end{matrix}\right) \,,
    \end{split}
    \label{eq_powerspec_recursion_BPP}
\end{equation}
where $c_x,c_y$ are the number of connections per predator and prey species respectively.
Analogous to Eq.~(\ref{eq_sr_meanfield}) we now derive a system of equations and solve for $r_x,r_y$ and $\phi_x,\phi_y$.
In the main text we describe the features of the power spectral density deduced from this system of equations.

\subsection*{Interpreting the Power Spectral Density in the Context of Temporal Stability}\label{sec_PSD_interp}

For orientation, we here provide some interpretation of the power spectral density in the context of temporal stability. Essentially when we talk about temporal stability, we can can be referring to one of two measures. The first is how far stochastic trajectories tend to stray from their equilibrium value over long time horizons. We refer to this as {`variability'}~\cite{donohue_dimensionality_2013}. The second is how quickly population abundances tend to change over finite time horizons. We will characterise this by the {`temporal autocorrelation'}.

\begin{figure}[t]
    \centering
    \includegraphics[width=0.48\textwidth]{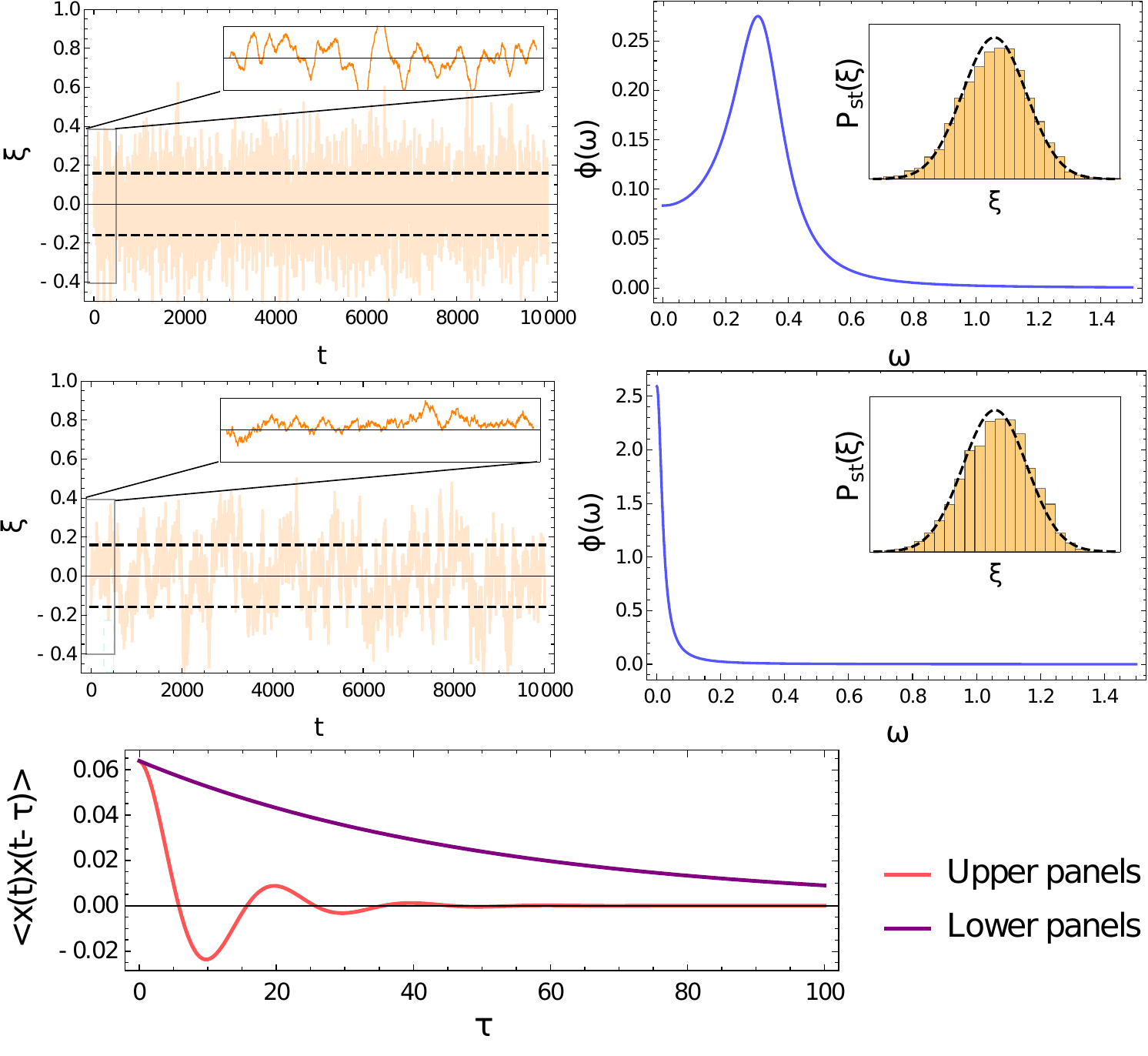}
    \caption{ 
{Example time series and corresponding fluctuation spectra for illustration.} Left panels: Examples of stochastic trajectories with $0$ means indicated by solid black lines and the standard deviations indicated by dashed lines. Insets show the same trajectories over a smaller time-window. Right panels: Corresponding power-spectral densities of the trajectories in the left hand panels. Insets show histograms of the corresponding stochastic trajectories, overlaid by the theoretical stationary distribution (black dashed line). Bottom panel: The temporal autocorrelation of the stochastic trajectories in the left hand panels can be obtained as the Fourier transform of the corresponding power spectra in the right hand panels.} 
    \label{fig_example_PS}
\end{figure}

The variability can be characterised by the variance in time-averaged trajectories around the mean~\cite{arnoldi_resilience_2016}. For a system such as Eq.~(\ref{eq2}), which we recall can be a linear approximation for a nonlinear system such as Eq.~(\ref{eq_HLV_sde_main}), we find that $\bm{\xi}$ is normally distributed with zero mean and a covariance matrix, $\bm{\Sigma}$, that solves the following Lyapunov equation~\cite{lyapunov};
\begin{equation}
 \bm{A} \bm{\Sigma} + \bm{\Sigma} \bm{A}^{T} + \bm{B}(\bm{x}^*) = 0 \,. \label{eq_lyapunov}
\end{equation}
The stationary distribution of $\bm{\xi}$ is then $P_{\mathrm{st}}(\bm{\xi})= \mathcal{N}(\bm{0},\bm{\Sigma})$. For instance, in the left panels of Fig.~\ref{fig_example_PS}, we show  stochastic trajectories for two different systems, with standard deviations marked by black dashed lines. Meanwhile the marginal normal distribution for these trajectories is plotted in the inset of the right panels of Fig.~\ref{fig_example_PS}. A system can then be said to be {`less stable'} (in a temporal sense) if it has a greater variability. A consideration of the solutions to Eq.~(\ref{eq_lyapunov}) shows that this measure of temporal stability is highly correlated with asymptotic stability; less stable deterministic systems tend to have stochastic counterparts with higher variance around equilibrium states.

Despite the fact that the trajectories in Fig.~\ref{fig_example_PS} have the same variance (see black dashed lines in left panels and inset plots in right panels) it is clear that they have very different temporal structure. While these differences are entirely masked by the measure of variability (which time-averages out the temporal structure), such differences are captured by the power-spectral density (see Fig.~\ref{fig_example_PS}, right panels). For instance, the peak at $\omega \approx 0.3$ in the power spectrum in the upper right hand panel indicates that the trajectories in the upper left panel exhibit quasi-cycles (i.e. have a typical frequency, see inset), while the peak at $\omega =0$ at following decay of the power spectrum in the middle right hand panel indicates that the trajectories in the middle left panel do not exhibit quasi-cycles (i.e. do not have a typical frequency, see inset).

In the context of temporal stability, the relationship between the power spectra and the autocorrelation function $\langle \xi(t) \xi(t-\tau) \rangle$ is of particular importance. By the Wiener–Khinchin theorem, we know that the autocorrelation function is given by the Fourier transform of the power spectrum. This is shown in the bottom panel of Fig.~\ref{fig_example_PS}. The autocorrelation of the trajectory in the upper panels decays rapidly with time. In contrast, the autocorrelation of the trajectory in the middle panel decays more slowly. This can be clearly seen in the inset trajectory plots (left hand panels, top and middle). Thus we see that a distinct measure of temporal stability exists that is more appropriate over shorter time horizons; a system can be said to be {`less stable'} over finite times if it has a more rapidly decaying autocorrelation function. This measure of temporal stability is more weakly correlated with asymptotic stability than its counterpart, variability, as it is affected by the magnitude imaginary parts of the system's eigenvalues (rather than their real parts, as in asymptotic stability).

\subsection*{Figure details}
\paragraph*{Figure 1, left panels}
A large random Lotka-Volterra ecosystem of the type described above was generated. Parameters used were: $N=1000, c=50, \gamma=-1, \sigma^2=1/4/c, b=0.05$. The solid line is the result of Eq.~(6), noting that $\mu=\sqrt{2\sigma^2/\pi}$. For the empirical power spectrum, we used an Euler-Maryama time-stepping method to simulate a time series of length $t_{\max}=2^{10}$ and time step $h=2^{-7}$. The power spectrum for each species was calculated with a Fast Fourier Transform, and the result averaged over all species. The top panel shows part of the time series generated for the first species. 
\paragraph*{Figure 1, right panel}
A two-trophic level model ecosystem was generated as described above. Parameters in this case were: $Nx=200, N_y=800, c_x=20, c_y=5, \alpha=10, b=1, d=1$. Time series and power spectra were computed similarly to the left panels. 
\paragraph*{Figure 2}
For the left panel, we generated Lotka-Volterra ecosystems with parameters $N=1000, c=50, \sigma^2=0.5$, using $b=2+(1+\gamma)\sqrt(c*\sigma^2)$ for the simulations with $\gamma=0$ and $\gamma=1$. Time series and spectra were computed similarly to Figure 1. For the right panels more care is needed. Finite random matrices typically have a small number of eigenvalues that are order $1/\sqrt{N}$ larger than predicted by the stability boundary in the limit $N\to\infty$. To achieve the near-instability results in this figure, we first generated the off-diagonal entries of the community matrices, then chose the birth rate $b$ to put the rightmost eigenvalue of $\bm{A}$ exactly at zero. 
\paragraph*{Figure 3} 
Parameters here are: $N=500, c=20, \gamma=-1, \sigma^2=1/4/c, b=0.2$. For the `direct' results we numerically computed the power spectral density according to the matrix formula in Eq.~(\ref{eq_powerspectrum_definition}). This was preferable to simulations of the time series, as a long time horizon is required to achieve good resolution of the individual contributions to the power spectral density. 
\paragraph*{Figure 4} 
Parameters here are: $N_x=100, N_y=200, c_x=20, c_y=10, \alpha=5, b=1, d=1$.
\paragraph*{Figure 5} \hspace{63mm}\\
The dataset 41467\_2017\_2571\_MOESM6\_ESM.xlsx was imported into Matlab and processed as follows: We took the average of the three reported daily measurements to construct an 88-day time series for each species. To limit boundary effects we discarded all species with at least one with zero measured abundance, in doing so retaining 100 species. The mean was subtracted and then the power spectrum fitted using the covariance method with 8th order autoregression. The model fitting was achieved with a non-linear least squares method applied to our equation (\ref{eq_sr_meanfield}), with parameters $b$, $c\sigma^2$ (a composite parameter), $\gamma$, and an additional scale parameter for overall noise strength.

\subsection*{Data Availability}
Plankton abundance data used in Fig~5 are taken from \cite{plankton_timeseries} available at:\\ \texttt{nature.com/articles/s41467-017-02571-4\#Sec24}. All simulation data can be reproduced using the code available at \texttt{http://doi.org/10.5281/zenodo.4720998}.

\subsection*{Code Availability}
Code to reproduce all Figures is available at \texttt{http://doi.org/10.5281/zenodo.4720998}.

\subsection*{Acknowledgements}
T.R. and Y.K. gratefully acknowledge the support of the Royal Society. G.W.A.C. thanks Leverhulme Trust for support through the Leverhulme Early Career Fellowship.

\subsection*{Author Contributions}
Y.K., Q.Y., G.W.A.C. and T.R. developed the theory and performed the computations. Y.K. wrote the manuscript with support from T.R. and G.W.A.C.. Y.K., Q.Y., G.W.A.C. and T.R. discussed and approved the final manuscript.

\subsection*{Competing Interests}
The authors declare no competing interests.

\bibliographystyle{unsrt}
\bibliography{references}

\begin{thebibliography}{10}

\bibitem{may_will_1972}
Robert~M. May.
\newblock Will a {Large} {Complex} {System} be {Stable}?
\newblock {\em Nature}, 238(5364):413, August 1972.

\bibitem{gardner_connectance_1970}
Mark~R. Gardner and W.~Ross Ashby.
\newblock Connectance of {Large} {Dynamic} ({Cybernetic}) {Systems}: {Critical}
  {Values} for {Stability}.
\newblock {\em Nature}, 228(5273):784--784, November 1970.

\bibitem{ginibre1965statistical}
Jean Ginibre.
\newblock Statistical ensembles of complex, quaternion, and real matrices.
\newblock {\em Journal of Mathematical Physics}, 6(3):440--449, 1965.

\bibitem{james_constructing_2015}
Alex James, Michael~J. Plank, Axel~G. Rossberg, Jonathan Beecham, Mark
  Emmerson, and Jonathan~W. Pitchford.
\newblock Constructing {Random} {Matrices} to {Represent} {Real} {Ecosystems}.
\newblock {\em The American Naturalist}, 185(5):680--692, May 2015.
\newblock Publisher: The University of Chicago Press.

\bibitem{jacquet2016no}
Claire Jacquet, Charlotte Moritz, Lyne Morissette, Pierre Legagneux,
  Fran{\c{c}}ois Massol, Philippe Archambault, and Dominique Gravel.
\newblock No complexity--stability relationship in empirical ecosystems.
\newblock {\em Nature communications}, 7:12573, 2016.

\bibitem{jr_unified_2010}
Egbert G.~Leigh Jr, James Rosindell, and Rampal~S. Etienne.
\newblock Unified neutral theory of biodiversity and biogeography.
\newblock {\em Scholarpedia}, 5(11):8822, November 2010.

\bibitem{allesina_stability_2012}
Stefano Allesina and Si~Tang.
\newblock Stability criteria for complex ecosystems.
\newblock {\em Nature}, 483(7388):205--208, March 2012.
\newblock Number: 7388 Publisher: Nature Publishing Group.

\bibitem{allesina_stabilitycomplexity_2015}
Stefano Allesina and Si~Tang.
\newblock The stability–complexity relationship at age 40: a random matrix
  perspective.
\newblock {\em Population Ecology}, 57(1):63--75, January 2015.
\newblock Publisher: John Wiley \& Sons, Ltd.

\bibitem{allesina2015predicting}
Stefano Allesina, Jacopo Grilli, Gy{\"o}rgy Barab{\'a}s, Si~Tang, Johnatan
  Aljadeff, and Amos Maritan.
\newblock Predicting the stability of large structured food webs.
\newblock {\em Nature communications}, 6(1):1--6, 2015.

\bibitem{grilli2016modularity}
Jacopo Grilli, Tim Rogers, and Stefano Allesina.
\newblock Modularity and stability in ecological communities.
\newblock {\em Nature communications}, 7(1):1--10, 2016.

\bibitem{donohue_2016}
Ian Donohue, Helmut Hillebrand, José~M. Montoya, Owen~L. Petchey, Stuart~L.
  Pimm, Mike~S. Fowler, Kevin Healy, Andrew~L. Jackson, Miguel Lurgi, Deirdre
  McClean, Nessa~E. O'Connor, Eoin~J. O'Gorman, and Qiang Yang.
\newblock Navigating the complexity of ecological stability.
\newblock {\em Ecology Letters}, 19(9):1172--1185, 2016.

\bibitem{grimm_babel_1997}
V.~Grimm and Christian Wissel.
\newblock Babel, or the ecological stability discussions: an inventory and
  analysis of terminology and a guide for avoiding confusion.
\newblock {\em Oecologia}, 109(3):323--334, February 1997.

\bibitem{levins_coexistence_1979}
Richard Levins.
\newblock Coexistence in a {Variable} {Environment}.
\newblock {\em The American Naturalist}, 114(6):765--783, December 1979.
\newblock Publisher: The University of Chicago Press.

\bibitem{ives_stability_1999}
A.~R. Ives, K.~Gross, and J.~L. Klug.
\newblock Stability and {Variability} in {Competitive} {Communities}.
\newblock {\em Science}, 286(5439):542--544, October 1999.

\bibitem{lehman_biodiversity_2000}
Clarence L. Lehman and David Tilman.
\newblock Biodiversity, {Stability}, and {Productivity} in {Competitive}
  {Communities}.
\newblock {\em The American Naturalist}, 156(5):534--552, November 2000.

\bibitem{tilman_biodiversity_2006}
David Tilman, Peter~B. Reich, and Johannes M.~H. Knops.
\newblock Biodiversity and ecosystem stability in a decade-long grassland
  experiment.
\newblock {\em Nature}, 441(7093):629--632, June 2006.

\bibitem{jiang_different_2009}
Lin Jiang and Zhichao Pu.
\newblock Different {Effects} of {Species} {Diversity} on {Temporal}
  {Stability} in {Single}‐{Trophic} and {Multitrophic} {Communities}.
\newblock {\em The American Naturalist}, 174(5):651--659, November 2009.

\bibitem{loreau_species_2008}
Michel Loreau and Claire de~Mazancourt.
\newblock Species {Synchrony} and {Its} {Drivers}: {Neutral} and {Nonneutral}
  {Community} {Dynamics} in {Fluctuating} {Environments}.
\newblock {\em The American Naturalist}, 172(2):E48--E66, August 2008.

\bibitem{campbell_experimental_2011}
Veronik Campbell, Grace Murphy, and Tamara~N. Romanuk.
\newblock Experimental design and the outcome and interpretation of
  diversity-stability relations.
\newblock {\em Oikos}, 120(3):399--408, March 2011.

\bibitem{donohue_dimensionality_2013}
Ian Donohue, Owen~L. Petchey, José~M. Montoya, Andrew~L. Jackson, Luke
  McNally, Mafalda Viana, Kevin Healy, Miguel Lurgi, Nessa~E. O'Connor, and
  Mark~C. Emmerson.
\newblock On the dimensionality of ecological stability.
\newblock {\em Ecology Letters}, 16(4):421--429, 2013.
\newblock \_eprint: https://onlinelibrary.wiley.com/doi/pdf/10.1111/ele.12086.

\bibitem{suweis2015effect}
Samir Suweis, Jacopo Grilli, Jayanth~R Banavar, Stefano Allesina, and Amos
  Maritan.
\newblock Effect of localization on the stability of mutualistic ecological
  networks.
\newblock {\em Nature communications}, 6(1):1--7, 2015.

\bibitem{arnoldi_resilience_2016}
J-F. Arnoldi, M.~Loreau, and B.~Haegeman.
\newblock Resilience, reactivity and variability: {A} mathematical comparison
  of ecological stability measures.
\newblock {\em Journal of Theoretical Biology}, 389:47--59, January 2016.

\bibitem{wiesenfeld1985noisy}
Kurt Wiesenfeld.
\newblock Noisy precursors of nonlinear instabilities.
\newblock {\em Journal of Statistical Physics}, 38(5-6):1071--1097, 1985.

\bibitem{kubo1966fluctuation}
Rep Kubo.
\newblock The fluctuation-dissipation theorem.
\newblock {\em Reports on progress in physics}, 29(1):255, 1996.

\bibitem{alonso2007stochastic}
David Alonso, Alan~J McKane, and Mercedes Pascual.
\newblock Stochastic amplification in epidemics.
\newblock {\em Journal of the Royal Society Interface}, 4(14):575--582, 2007.

\bibitem{galla2009intrinsic}
Tobias Galla.
\newblock Intrinsic noise in game dynamical learning.
\newblock {\em Physical review letters}, 103(19):198702, 2009.

\bibitem{mckane_predator-prey_2005}
A.~J. McKane and T.~J. Newman.
\newblock Predator-{Prey} {Cycles} from {Resonant} {Amplification} of
  {Demographic} {Stochasticity}.
\newblock {\em Physical Review Letters}, 94(21):218102, June 2005.

\bibitem{wigner1958distribution}
Eugene~P Wigner.
\newblock On the distribution of the roots of certain symmetric matrices.
\newblock {\em Annals of Mathematics}, pages 325--327, 1958.

\bibitem{tao2008random}
Terence Tao and Van Vu.
\newblock Random matrices: the circular law.
\newblock {\em Communications in Contemporary Mathematics}, 10(02):261--307,
  2008.

\bibitem{tao2010random}
Terence Tao, Van Vu, Manjunath Krishnapur, et~al.
\newblock Random matrices: Universality of esds and the circular law.
\newblock {\em The Annals of Probability}, 38(5):2023--2065, 2010.

\bibitem{stone_feasibility_2018}
Lewi Stone.
\newblock The feasibility and stability of large complex biological networks: a
  random matrix approach.
\newblock {\em Scientific Reports}, 8(1):8246, May 2018.
\newblock Number: 1 Publisher: Nature Publishing Group.

\bibitem{gibbs_effect_2018}
Theo Gibbs, Jacopo Grilli, Tim Rogers, and Stefano Allesina.
\newblock Effect of population abundances on the stability of large random
  ecosystems.
\newblock {\em Physical Review E}, 98(2):022410, August 2018.

\bibitem{gardiner_stochastic_2009}
Crispin Gardiner.
\newblock {\em Stochastic {Methods}: {A} {Handbook} for the {Natural} and
  {Social} {Sciences}}.
\newblock Springer {Series} in {Synergetics}. Springer-Verlag, Berlin
  Heidelberg, 4 edition, 2009.

\bibitem{plankton_timeseries}
Antonio~M. Martin-Platero, Brian Cleary, Kathryn Kauffman, Sarah~P. Preheim,
  Dennis~J. McGillicuddy, Eric~J. Alm, and Martin~F. Polz.
\newblock High resolution time series reveals cohesive but short-lived
  communities in coastal plankton.
\newblock {\em Nature Communications}, 9:266, 2018.

\bibitem{coyte_2015}
Katharine~Z. Coyte, Jonas Schluter, and Kevin~R. Foster.
\newblock The ecology of the microbiome: Networks, competition, and stability.
\newblock {\em Science}, 350(6261):663--666, 2015.

\bibitem{stein_2013}
Richard~R. Stein, Vanni Bucci, Nora~C. Toussaint, Charlie~G. Buffie, Gunnar
  Rätsch, Eric~G. Pamer, Chris Sander, and João~B. Xavier.
\newblock Ecological modeling from time-series inference: Insight into dynamics
  and stability of intestinal microbiota.
\newblock {\em PLOS Computational Biology}, 9(12):1--11, 12 2013.

\bibitem{xiao_2017}
Yandong Xiao, Marco~Tulio Angulo, Jonathan Friedman, Matthew~K. Waldor,
  Scott~T. Weiss, and Yang-Yu Liu.
\newblock Mapping the ecological networks of microbial communities.
\newblock {\em Nature Communications}, 8:2042, 2017.

\bibitem{galla2018dynamically}
Tobias Galla.
\newblock Dynamically evolved community size and stability of random
  lotka-volterra ecosystems (a).
\newblock {\em EPL (Europhysics Letters)}, 123(4):48004, 2018.

\bibitem{gibbs2018effect}
Theo Gibbs, Jacopo Grilli, Tim Rogers, and Stefano Allesina.
\newblock Effect of population abundances on the stability of large random
  ecosystems.
\newblock {\em Physical Review E}, 98(2):022410, 2018.

\bibitem{mckane2014stochastic}
Alan~J McKane, Tommaso Biancalani, and Tim Rogers.
\newblock Stochastic pattern formation and spontaneous polarisation: the linear
  noise approximation and beyond.
\newblock {\em Bulletin of mathematical biology}, 76(4):895--921, 2014.

\bibitem{picoche2020}
Coralie Picoche and Frédéric Barraquand.
\newblock Strong self-regulation and widespread facilitative interactions in
  phytoplankton communities.
\newblock {\em Journal of Ecology}, 108(6):2232--2242, 2020.

\bibitem{pennington2017nonlinear}
Jeffrey Pennington and Pratik Worah.
\newblock Nonlinear random matrix theory for deep learning.
\newblock In {\em Advances in Neural Information Processing Systems}, pages
  2637--2646, 2017.

\bibitem{moran2019will}
Jos{\'e} Moran and Jean-Philippe Bouchaud.
\newblock Will a large economy be stable.
\newblock {\em Available at SSRN}, 2019.

\bibitem{luo2007constructing}
Feng Luo, Yunfeng Yang, Jianxin Zhong, Haichun Gao, Latifur Khan, Dorothea~K
  Thompson, and Jizhong Zhou.
\newblock Constructing gene co-expression networks and predicting functions of
  unknown genes by random matrix theory.
\newblock {\em BMC bioinformatics}, 8(1):299, 2007.

\bibitem{almog2019uncovering}
Assaf Almog, M~Renate Buijink, Ori Roethler, Stephan Michel, Johanna~H Meijer,
  Jos~HT Rohling, and Diego Garlaschelli.
\newblock Uncovering functional signature in neural systems via random matrix
  theory.
\newblock {\em PLoS computational biology}, 15(5):e1006934, 2019.

\bibitem{stratonovich1957}
R.~L. {Stratonovich}.
\newblock {On a Method of Calculating Quantum Distribution Functions}.
\newblock {\em Soviet Physics Doklady}, 2:416, July 1957.

\bibitem{hubbard1959}
J.~Hubbard.
\newblock Calculation of partition functions.
\newblock {\em Phys. Rev. Lett.}, 3:77--78, Jul 1959.

\bibitem{mezard_sk_1986}
M~Mézard, G~Parisi, and M.~A Virasoro.
\newblock {SK} {Model}: {The} {Replica} {Solution} without {Replicas}.
\newblock {\em Europhysics Letters (EPL)}, 1(2):77--82, January 1986.

\bibitem{mezard_spin_1987}
Marc Mezard, Giorgio Parisi, and Miguel~Angel Virasoro.
\newblock {\em Spin glass theory and beyond}.
\newblock Number v. 9 in World {Scientific} lecture notes in physics. World
  Scientific, Singapore ; New Jersey, 1987.
\newblock OCLC: ocm14929802.

\bibitem{rogers_cavity_2008}
Tim Rogers, Isaac~Pérez Castillo, Reimer Kühn, and Koujin Takeda.
\newblock Cavity approach to the spectral density of sparse symmetric random
  matrices.
\newblock {\em Physical Review E}, 78(3):031116, September 2008.

\bibitem{rogers_cavity_2009}
Tim Rogers and Isaac~Pérez Castillo.
\newblock Cavity approach to the spectral density of non-{Hermitian} sparse
  matrices.
\newblock {\em Physical Review E}, 79(1):012101, January 2009.

\bibitem{metz2019spectral}
Fernando~Lucas Metz, Izaak Neri, and Tim Rogers.
\newblock Spectral theory of sparse non-hermitian random matrices.
\newblock {\em Journal of Physics A: Mathematical and Theoretical},
  52(43):434003, 2019.

\bibitem{lyapunov}
R.~A. Horn and C.~R. Johnson.
\newblock {\em {T}opics in {M}atrix {A}nalysis}.
\newblock Cambridge University Press, Cambridge, 1991.

\end{thebibliography}

\end{document}